\documentclass[aps,pre,superscriptaddress,amsmath,amssymb]{revtex4-1}

\usepackage{graphicx,xcolor}

\begin{document}


\title{Absolute/convective secondary instabilities and the role of confinement in free shear layers}


\author{Crist\'obal Arratia}
\email[]{cristobal.arratia@gmail.com}
\affiliation{Nordita
KTH Royal Institute of Technology and Stockholm University
Roslagstullsbacken 23, SE-106 91 Stockholm, Sweden}
\affiliation{Laboratory of Fluid Mechanics and Instabilities, EPFL, 1015 Lausanne, Switzerland}
\author{Saviz Mowlavi}
\affiliation{Laboratory of Fluid Mechanics and Instabilities, EPFL, 1015 Lausanne, Switzerland}
\affiliation{Department of Mechanical Engineering, Massachusetts Institute of Technology, Cambridge MA 02139, USA} 
\author{Fran\c cois Gallaire}
\affiliation{Laboratory of Fluid Mechanics and Instabilities, EPFL, 1015 Lausanne, Switzerland}

\date{\today}

\begin{abstract}
We study the linear spatiotemporal stability of an infinite row of equal point-vortices under symmetric confinement between parallel walls. 
These rows of vortices serve to model the secondary instability leading to the merging of consecutive (Kelvin-Helmholtz) vortices in free shear layers, allowing us to study how confinement limits the growth of shear layers through vortex pairings. 
Using a geometric construction akin to a Legendre transform on the dispersion relation, we compute the growth rate of the instability in different reference frames as a function of the frame velocity with respect to the vortices. 
This novel approach is verified and complemented with numerical computations of the linear impulse response, fully characterizing the absolute/convective nature of the instability. 
Similar to results by Healey on the primary instability of parallel tanh profiles [J. Fluid Mech. {\bf 623}, 241 (2009)], we observe a range of confinement in which absolute instability is promoted. 
For a given set of shear layer and channel width of a parallel shear layer, the threshold for absolute/convective instability of the secondary pairing instability depends on the separation distance between consecutive vortices, which is physically determined by the wavelength selected by the previous (primary or pairing) instability.  
In the presence of counterflow and moderate to weak confinement, small wavelength of the vortex row leads to absolute instability. 
We comment on the general significance of such absolute secondary instabilities, which have been previously related to an abrupt emergence of complex behaviour. 
We argue that in the present case, in which the result of the secondary pairing instability is to regenerate the flow with an increased wavelength eventually leading to convective instability, a different interpretation is possible. 
According to this new interpretation, a spatially developing row of vortices in a free shear layer with counterflow can only occur if the pairing instability is convective. 
This corresponds to a wavelength selection criteria, according to which the distance between consecutive vortices should be sufficiently large in comparison to the channel width. 
We argue that the proposed wavelength selection mechanism can serve as a guideline for experimentally obtaining plane shear layers with counterflow, which has remained an experimental challenge.

\end{abstract}


\maketitle

\section{Introduction}
Free shear or mixing layers form when two volumes of fluid travelling parallel to each other at different speeds are put in contact.
They appear in a large variety of natural and technological contexts. In addition, they constitute a basic building block for understanding other flows such as wakes, jets, or separation regions, since they provide the most basic example of shear flow instability: the instability of an inflection point in the velocity profile violating Rayleigh's criterion for stability of inviscid plane flow~\cite{DrazinReid2004}.
Known as Kelvin-Helmholtz (KH) instability, it leads to the roll-up of the vorticity of the shear layer into a row of consecutive vortices of the same sign~\cite{brown1974,hoHuerr1984}, the KH vortices, often referred to as billows or rollers.

A large body of work has been dedicated to the instabilities of mixing layers, 
which can be broadly divided into studies concerned with (i) the primary instability of the parallel or weakly non-parallel flow fields~\cite{Michalke1964,huerrmonk85,healey2009,gallaire2015}, and (ii) secondary instabilities of the already formed KH vortices~\cite{winant1974,PierrehumbertWidnallJFM1982,PeltierCaulfieldAnnRevFM2003,ACC2013}, which form later in time or further downstream. 
Several analysis have also included the effects of varying fluid properties like density or viscosity, but we will only consider the case of homogeneous shear layers in which the two flow streams are composed of the same fluid. 

In this context, the main aspect of the primary instability (i) that has remained a subject of research during the last few decades concerns the spatio-temporal stability~\cite{huerrmonk85,healey2009,gallaire2015}. 
This sprung from the seminal recognition by Huerre \& Monkewitz\cite{huerrmonk85} of the importance of the distinction between absolute and convective instabilities~\cite{huerrMonk1990}, previously developed in plasma physics~\cite{briggs1964}. 
This theory provided an excellent prediction of the experimentally observed transition threshold from a band-pass noise-amplifier to a self-sustained peaked oscillator as the counterflow is increased~\cite{StrykowskiNiccum1991}, that is, as the mean velocity of the two streams is reduced while maintaining the same amount of shear. 
This provided the first experimental confirmation of the fundamental role of the absolute/convective nature in the spectral signature of instabilities in open flows. 
More recently, Healey~\cite{healey2009} demonstrated that a moderate amount of confinement in the shearing direction promotes the absolute nature of the instability, without however providing any experimental evidence. 
But as remarked 
previously by Juniper~\cite{Juniper2006ConfinementEffect,Juniper2007FullResponseAndConfinement}, a similar destabilizing effect of confinement is present in jets and wakes, for which there are indeed experimental indications of such an effect~\citep[see][and references therein]{Juniper2006ConfinementEffect}. 
Unsurprisingly, these flows are all stabilized for sufficiently strong confinement. 

Studies of secondary instabilities (ii) have uncovered several physical mechanisms, such as for instance the elliptic and hyperbolic instabilities, which break the translational invariance in the spanwise direction through three dimensional (3D) instabilities~\cite{PierrehumbertWidnallJFM1982,PeltierCaulfieldAnnRevFM2003,ACC2013}. 
A seminal contribution here is the study of Pierrehumbert \& Widnall~\cite{PierrehumbertWidnallJFM1982} on the 3D stability of Stuart vortices~\cite{stuart1967}. 
These vortices are given by a one-parameter family of 2D solutions to the Euler equations proposed to represent a periodic array of 2D vortices separating two counter-flows, with a parameter $\rho \in [0,1]$ measuring the dimensionless vortex core size. 
The solution for $\rho=1$ consists of a periodic row of infinitely concentrated point vortices, while $\rho=0$ corresponds to the hyperbolic-tangent parallel velocity profile. 
We focus our attention here on the two-dimensional (2D) pairing instability associated with the growth of the first subharmonic~\cite{saffman1992}. 
Eventually, this secondary instability saturates and leads to the merger of primary vortices into larger vortices with twice the initial spacing~\cite{winant1974}. 

Since it leads to a similar row of vortices as that from which it develops, this secondary instability can then repeat itself in a sequence of successive instabilities, which can develop in time or in space.   
The first case is the temporal shear layer, as in numerical simulations~\cite{CorcosShermanJFM1984MixingLayerPart1} and tilt-tank experiments~\cite{thorpe_1973}, in which there is no mean advection and each (numerical or physical) experiment is a one time transient. 
The second case, that of spatial mixing layers, is the most thoroughly studied experimentally~\cite{hoHuerr1984}, most often as co-flowing mixing layers. 
These experiments can in principle last for an arbitrarily large amount of time, during which subsequent instabilities succeed one after another in different and more or less stationary regions of space. 
In this context, the pairing instability was described by Winant \& Browand~\cite{winant1974} as the key mechanism of spatial growth for turbulent mixing layers. 
In the absence of forcing, they report a large variability on the pairing locations and observed up to four pairings~\cite{winant1974}. 
In experiments on an axisymmetric air jet at $Re=50000$ with an applied periodic forcing, Kibens~\cite{kibens1980pairings} identified a sequence of three successive vortex-pairing events at well fixed downstream locations. 
This large variability and high receptivity to forcing are now well understood as the footprint of the band-pass noise-amplifier behaviour of convective instabilities. 
It is interesting to note that the previously mentioned experimental confirmation of a transition from a band-pass noise-amplifier to a self-sustained peaked oscillator with increasing counterflow was achieved on a spatial axisymmetric air jet~\cite{StrykowskiNiccum1991}. 
An earlier attempt at generating a spatial mixing layer with counterflow failed, resulting instead in a flow with a totally different configuration consisting roughly in a single stagnation point with hyperbolic streamlines~\cite{humphreyLi1981tilting}. 
A planar mixing layer with counterflow was achieved much later with a more elaborate confining geometry~\cite{ForlitiEtal2005}. 

As stated above, the distinction between absolute and convective instabilities has been found determinant in the behaviour of many unstable open flows, but the vast majority of studies have focused on its effect on the primary instability. 
The absolute/convective nature of secondary instabilities have received far less attention, although they can grow not only in time but also in space, just as primary disturbances. 
In the words of Huerre~\cite{Huerre1988}, ``primary and  secondary instabilities arising in fluid flows need not have the same absolute/convective character". 
Some absolute/convective analysis of secondary instabilities include those of the Ekhaus and zig-zag instabilities \cite{ChomazCouaironJulien1999}, the subharmonic instability of a periodic array of vortex rings \cite{bolnotLedizesLewekeFDR2014} or the von Karman street of alternating point vortices \cite{mowlavi2016}. 
There are no major difficulties when the secondary instability is convective. 
Indeed, causality implies that a temporal sequence of instabilities can translate into a spatial one if the instabilities are convective~\cite{brancher1997}. 
In this case the secondary instability may develop downstream on the saturated state without affecting its precursor, the primary instability upstream. 
We have previously shown~\cite{mowlavi2016} that this is the case for the K\'arm\'an street of point vortices~\cite{lamb1932} modelling vortex shedding phenomena: the well known (secondary) instability~\cite{saffman1992} was shown to be strongly convective when applied to wakes~\cite{mowlavi2016}, thus reconciling the intrinsic instability of K\'arm\'an's point-vortex model with ubiquitous observations of vortex shedding behind obstacles. 
It has been recently confirmed experimentally~\cite{BonifaceEtalEPL2017} that this instability can be also stabilized by strong confinement, even in the absence of mean advection. 

The situation is less straightforward if the secondary instability is absolute. 
From a certain perspective this can be seen as somewhat paradoxical, since an instability propagating upstream would disrupt its precursor~\cite{brancher1997}. 
This situation has been studied by Chomaz and coworkers\cite{brancher1997,CouaironChomaz1ary2dary1999,ChomazCouaironJulien1999}, who have proposed that it constitutes a possible scenario for abrupt transition~\cite{ChomazARFM2005,Chomaz2004EJMBSecondaryAbsoluteAA}. 
According to Chomaz' description, if the secondary instability is already absolute when the primary instability transitions from convective to absolute, perturbations in the lee of the primary front never fade away and are likely to yield a complex, disordered behaviour. 
An example of this was observed by Couairon \& Chomaz in a complex Ginzburg-Landau model~\cite{CouaironChomaz1ary2dary1999}, in which the system presented aperiodic behaviour when the primary nonlinear wave was subject to absolute secondary instability. 
Moreover, in the words of his review~\cite[][p. 385]{ChomazARFM2005}, this 
``one-step scenario to disorder, involving a global mode made of a wave already absolutely unstable to secondary instability at the global threshold, may also explain the abrupt transition to turbulence observed in the rotating disk..." 
Chomaz then refers to theoretical and experimental studies by Lingwood, showing that turbulence in rotating disk flow occurs close to where the flow becomes absolutely unstable~\cite{lingwood_1995,lingwood1996experimental}, and work by Pier showing that there is an absolute secondary instability in this flow and proposing it as the mechanism of direct transition~\cite{Pier2003}. 
Recent investigations~\cite{viaud_serre_chomaz_2011,imayama_alfredsson_lingwood_2014,appelquist_schlatter_alfredsson_lingwood_2016} yield support to this scenario, 
although the secondary instability is not the subharmonic one predicted by Pier~\cite{Pier2003} and its nature is not yet clear. 

Returning to free shear layers, Brancher \& Chomaz~\cite{brancher1997} have indeed shown that the secondary pairing instability is more prompt to become absolute than the primary one. 
More precisely, they have determined the absolute/convective nature of the subharmonic pairing instability of a row of finite-size co-rotating vortices, viewed as the saturation of the primary Kelvin-Helmhotz instability. 
For that purpose, they used the family of Stuart vortices and performed a spatiotemporal stability analysis through the numerical calculation of the linearized impulse response, thereby generalizing the temporal stability analysis of Ref.~\cite{PierrehumbertWidnallJFM1982}. 
They found that the backflow needed to trigger absolute instability was monotonically decreasing when the vortex concentration was increasing. 
In particular, the secondary pairing instability was found already absolutely unstable for backflows for which the primary instability was still convective. 
According to Chomaz~\cite[][p. 384]{ChomazARFM2005}, ``[t]hese results on the pairing instability explain why the 2D Global mode in the parallel mixing layer computed \citep[in Ref.][]{Chomaz2004EJMBSecondaryAbsoluteAA} is irregular at threshold, with pairings occurring randomly."
We will come back to this quotation and argue that a different interpretation is also possible in this case. 
In particular, the effects of confinement have not been explored, and as the spacing between consecutive vortices increases, these should eventually become important. 
{\it The near self-similarity of the sequential process of mixing layer growth through vortex pairings is most naturally broken by the external length-scale imposed by confinement.} 
As we shall see, consideration of confinement provides a mechanism for wavelength selection in spatially developing shear layers with counterflow. 

In the present study, we focus on the effect of confinement on the absolute/convective nature of the secondary instability of a periodic row of point vortices. 
In Sec.~\ref{sec:ProblemFormulation}, we describe our model and formulate the linear stability problem yielding the relevant dispersion relation. 
We compute the growth rate of the instability in different reference frames in Sec.~\ref{sec:spatiotemporal}. 
For this, we develop in Sec.~\ref{sec:LegendreGeoConstruction} a geometrical method consisting essentially of a Legendre transformation that effectively switches the dependence on the imaginary part of the complex wavenumber by a dependence on the propagation velocity of a moving frame. 
A more challenging application of this new method is given for the K\'arm\'an street of point vortices in Appendix~\ref{ap:karmanLegendre}. 
In Sec.~\ref{sec:applications} we provide a comparison with similar spatiotemporal stability results for a parallel $\tanh$ profile and discuss the possible implications of our results, including a wavelength selection mechanism that provides a criteria for the admissibility of plane shear layers (in the form of vortex rows) with counterflow. We summarize and conclude in Sec.~\ref{sec:conclusion} 

\section{Problem formulation}
\label{sec:ProblemFormulation}

\subsection{Governing equations}

\begin{figure}
\includegraphics[width=0.5\textwidth]{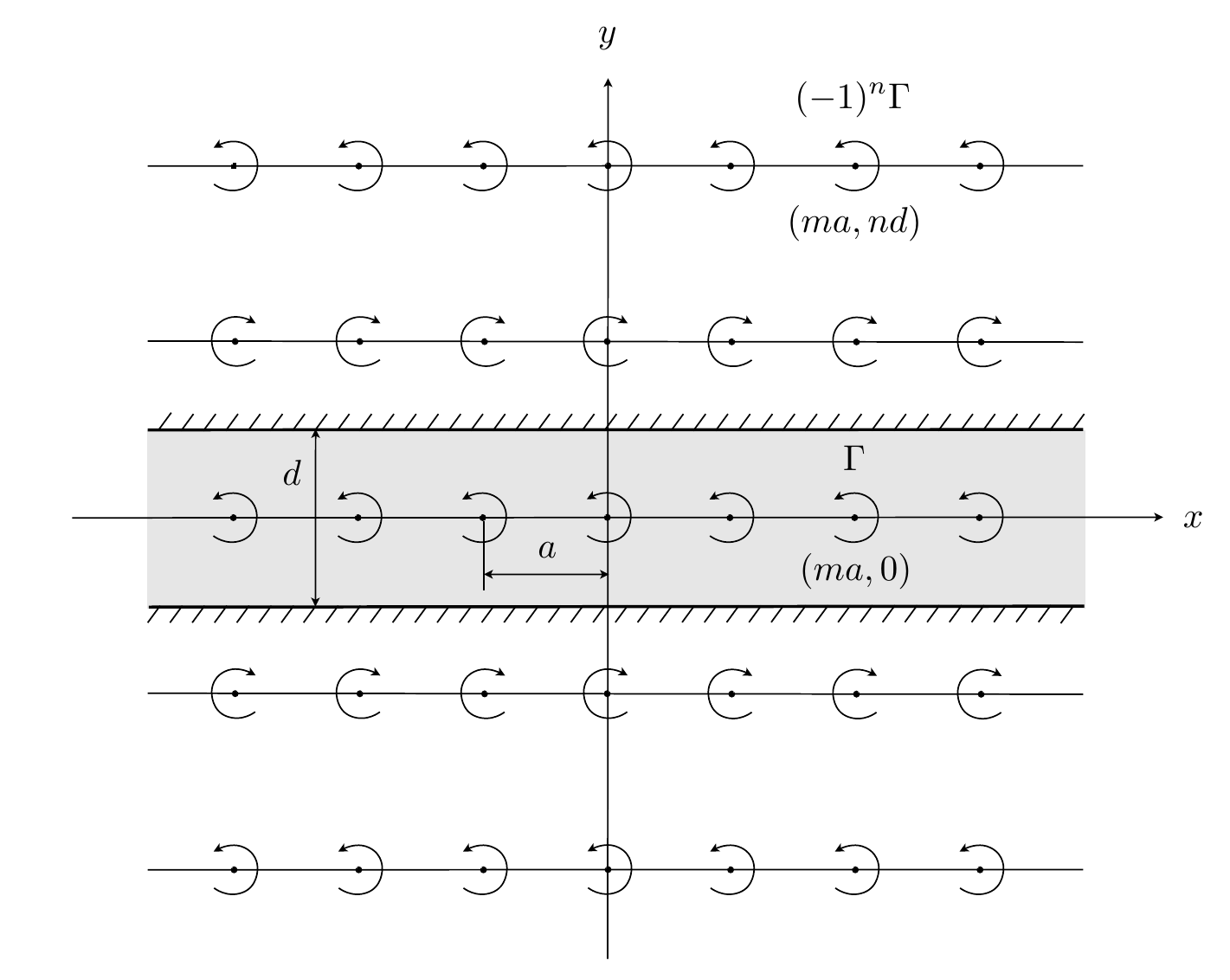}
\caption{Confined single row of point vortices.}
\label{fig:ConfinedSingleRow}
\end{figure}

We consider the system shown in Fig. \ref{fig:ConfinedSingleRow}, composed of an infinite row of point vortices of circulation $\Gamma$ symmetrically enclosed between two horizontal confining walls.
The distance between consecutive vortices is $a$ and that between the two walls is $d$.
The confinement is imposed by assigning to each vortex an infinite series of image vortices above and below the walls. 
This gives rise to a doubly infinite array of vortices situated at coordinates $(ma,nd)$ and of strength $(-1)^n \Gamma$, where $m,n \in \mathbb{Z}$ and $n = 0$ corresponds to the physical vortices.

This array of vortices corresponds to a static equilibrium configuration since the velocity induced on any of the physical vortices by all other vortices is zero. 
Applying a perturbation $(x_m,y_m)$ to the positions of the physical vortices, the coordinates of the vortices become $(ma + x_m, nd + (-1)^n y_m),$ wherein all the image vortices are duly tied to their corresponding physical vortex. Without loss of generality, the equations of motion of the $m = 0$ physical vortex are
\begin{subequations}
\begin{align}
\frac{dx_0}{dt} & = - a^2\omega_0
\sum_{m = -\infty}^{\infty} \ \sideset{}{''}\sum_{n = -\infty}^{\infty} (-1)^n \frac{y_0-nd-(-1)^n y_m}{r_{m,n}^2}, \\ 
\frac{dy_0}{dt} & = a^2\omega_0
\sum_{m = -\infty}^{\infty} \ \sideset{}{''}\sum_{n = -\infty}^{\infty} (-1)^n \frac{x_0-ma-x_m}{r_{m,n}^2},
\end{align}
\label{eq:EquationsMotion}
\end{subequations}
where 
\begin{equation}
r_{m,n}^2 = (nd+(-1)^n y_m-y_0)^2 + (ma+x_m-x_0)^2,
\end{equation}
is the squared distance between the $m = 0$ physical vortex and the $(m,n)$-vortex, and $\omega_0 = \Gamma/(2 \pi a^2)$. The double prime on the summation sign means that $n = 0$ is excluded when $m = 0.$ In the following, all quantities are nondimensionalized with the length $a$ and the characteristic time $1/\omega_0.$ 
The system is entirely specified by a single parameter, the confinement ratio $q = d/a$.

Assuming infinitesimal perturbations, Eqs.\,\eqref{eq:EquationsMotion} can be linearized about the equilibrium configuration to yield
\begin{subequations}
 \begin{align}
   \frac{dx_0}{dt} & = - \frac{\pi^2}{2q^2} y_0 - \sideset{}{'}\sum_{m = -\infty}^{\infty} \sum_{n = -\infty}^{\infty} \frac{n^2 q^2-m^2}{(n^2 q^2+m^2)^2} (y_m-(-1)^n y_0),\label{eq:linEqx} \\
  \frac{dy_0}{dt} & = \sideset{}{'}\sum_{m = -\infty}^{\infty} \sum_{n = -\infty}^{\infty} (-1)^n \frac{-n^2 q^2+m^2}{(n^2 q^2+m^2)^2} (x_m-x_0),
 \end{align}
\end{subequations}
where the prime on the summation sign means $m = 0$ is excluded. 
The series of images of the $m=0$ vortex yields the first term in 
Eq.\,\eqref{eq:linEqx}. 
Summing the series over $n$ in the software Mathematica gives
\begin{subequations}
 \begin{align}
 \frac{dx_0}{dt} & = \frac{\pi^2}{q^2} \sideset{}{'}\sum_{m = -\infty}^{\infty} C_m y_m - \frac{\pi^2}{q^2}\left[\frac{1}{2}+ \sideset{}{'}\sum_{m = -\infty}^{\infty} D_m \right] y_0 \\
 \frac{dy_0}{dt} & = \frac{\pi^2}{q^2} \sideset{}{'}\sum_{m = -\infty}^{\infty} D_m (x_m-x_0).
 \end{align}
\label{eq:LinearizedEquationsMotion}
\end{subequations}
where
\begin{subequations}\label{eq:defSeriesTerms}
 \begin{equation}
  C_m = \mathrm{csch}^2 \left( \frac{m\pi}{q} \right), 
 \end{equation}
 and
 \begin{equation}
  D_m = \frac{1}{4}\left[\mathrm{csch}^2 \left( \frac{m\pi}{2q} \right) + \mathrm{sech}^2 \left( \frac{m\pi}{2q} \right)\right].
 \end{equation}
\end{subequations}
These equations, defined here for the $m=0$ vortex, apply to each vortex $m$ and define the infinite set of linear equations governing the evolution of infinitesimal perturbations to the confined single row of point vortices.

\subsection{Dispersion relation and temporal stability} \label{sec:TemporalStability}

Let us now look for solutions to the perturbation Eqs.\,\eqref{eq:LinearizedEquationsMotion} of the form
\begin{equation}
\left[
\begin{array}{c}
x_m \\
y_m
\end{array}
\right] = \left[
\begin{array}{c}
\alpha \\
\beta 
\end{array}
\right]
e^{i(km-\omega t)},
\label{eq:Perturbation}
\end{equation}
where $k$ and $\omega$ are the wavenumber and frequency, respectively. 
Introducing \eqref{eq:Perturbation} into \eqref{eq:LinearizedEquationsMotion}, the governing equations are reduced to two coupled equations for $\alpha$ and $\beta$
\begin{equation}\label{eq:matrixEq}
\left[
\begin{array}{cc}
i \omega & A \\
B & i \omega
\end{array}
\right]
\left[
\begin{array}{c}
\alpha \\
\beta
\end{array}
\right] = \left[
\begin{array}{c}
0 \\
0
\end{array}
\right],
\end{equation}
where the coefficients $A$ and $B$ are given by 
\begin{subequations}\label{eq:defABinDispRel}
\begin{align}
A & =  \frac{\pi^2}{q^2} \Bigg[ \ \sideset{}{'}\sum_{m = -\infty}^{\infty} C_m e^{ikm} - \frac{1}{2} - \sideset{}{'}\sum_{m = -\infty}^{\infty} D_m \Bigg], \\
B & =  \frac{\pi^2}{q^2} \sideset{}{'}\sum_{m = -\infty}^{\infty} D_m (e^{ikm}-1),
\end{align}
\end{subequations}
with $C_m$ and $D_m$ defined in Eqs.\,\eqref{eq:defSeriesTerms}. The existence of non-trivial solutions to Eq.\,\eqref{eq:matrixEq} requires the determinant of the matrix to be zero, imposing the dispersion relation
\begin{equation}
\omega = \pm i \sqrt{AB}.
\label{eq:DispersionRelation}
\end{equation}

The temporal stability of the confined single row of vortices is determined by assigning a real value to the perturbation wavenumber $k$ and evaluating the growth rate of the perturbation, given by the imaginary part of $\omega$. We restrict ourselves to values of $k$ between $0$ and $\pi$ since disturbances of wavenumber $2\pi - k$ are equivalent to the complex conjugate of disturbances of wavenumber $k$. 

\begin{figure}
\includegraphics[width=0.5\textwidth]{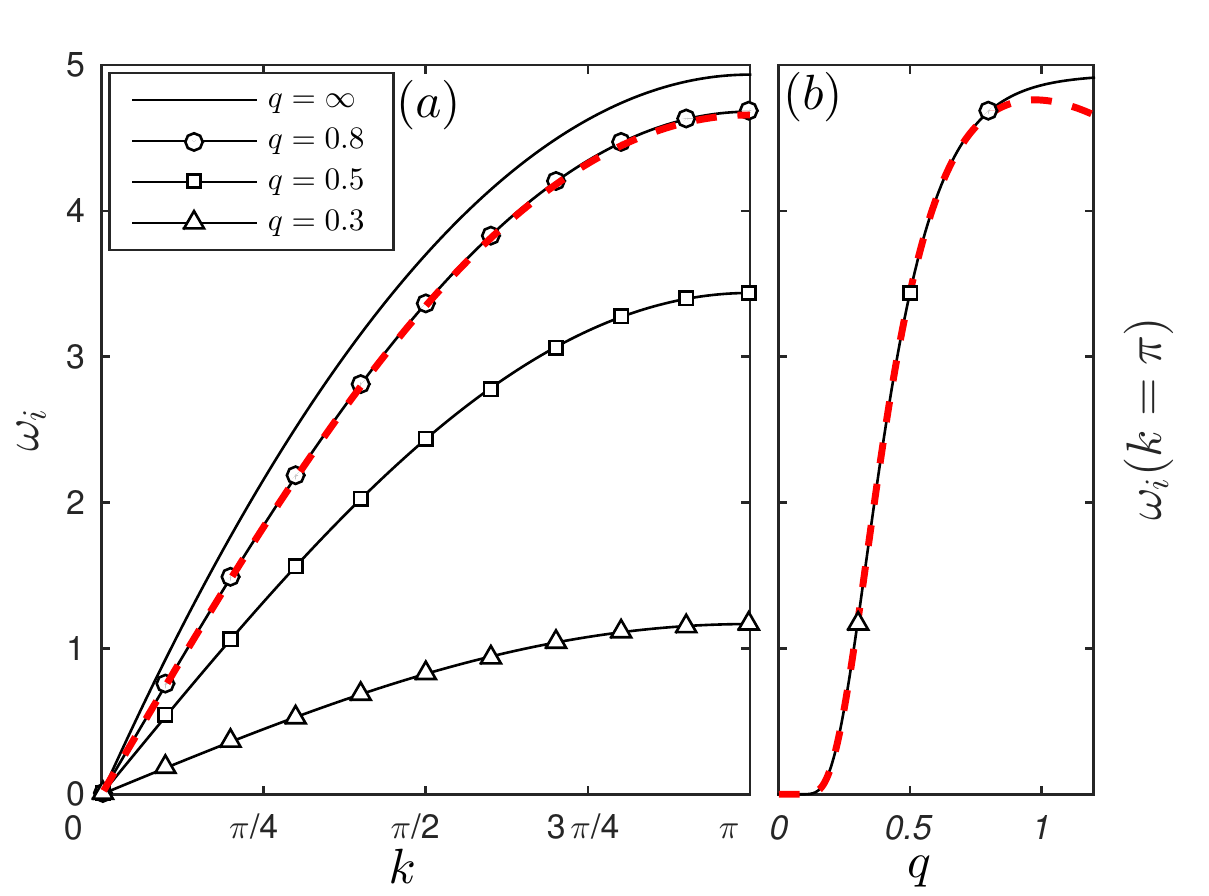}
\caption{Temporal dispersion relation. (a) Growth rate $\omega_i$ versus wavenumber $k$ for different confinement ratios. (b) Its value at $k = \pi$ versus confinement ratio $q$.}
\label{fig:TemporalStability}
\end{figure}

In Fig.~\ref{fig:TemporalStability}(a), we plot the growth rate $\omega_i$ versus the wavenumber $k$, for different values of the confinement ratio $q$. 
In all cases, the growth rate has a maximum at $k = \pi$ which corresponds to the vortex pairing instability \cite{saffman1992,winant1974}. 
While the growth rate decreases with the confinement ratio $q$, the instability is not totally supressed. 
Indeed, the product $AB$ is always a positive real number. 
Thus, the two solutions for $\omega$ are purely imaginary complex conjugates and the confined single row of vortices is temporally unstable for all finite values of $q$. 
This implies that the pairing instability is present regardless of how far appart along the channel are consecutive vortices, which is somewhat surprising, 
However, a simple analysis of the dispersion relation shows that the growth rate of the instability quickly goes to zero as $q\rightarrow 0.$ 
For small $q,$ each of the series coefficients $C_m$ and $D_m$ (defined in Eqs. \eqref{eq:defSeriesTerms}) go to zero exponentially and the series is dominated by the very first terms. 
Truncating the series while leaving the first two dominant terms with $k$ dependence gives that the dispersion relation goes as
\begin{equation}
\omega_i \sim \pm\frac{\pi^2}{q^2}  \exp\left(-\frac{\pi}{2q}\right) \sqrt{2(1-\cos k)(1 + 8 e^{-\pi/q})+4\sin^2(k)\,e^{-\pi/q}},
\label{eq:limitqto0DispersionRelation}
\end{equation}
when $q\rightarrow 0,$ showing that the growth rate $\omega_i$ is exponentially small for strong confinement. 
As shown in the dashed lines of figure \ref{fig:TemporalStability}(a), Eq. (\ref{eq:limitqto0DispersionRelation}) matches very well the temporal dispersion relation for a confinement ratio as large as $q=0.8$, which is already quite close to the dispersion relation of the unconfined case $q=\infty$. 

The unconfined limit is attained remarkably fast, as can be seen from the maximum growth rate $\omega_i(k=\pi)$ as a function of $q$ shown in figure \ref{fig:TemporalStability}(b). 
The maximum $\omega_i$ for $q=1.2,$ at the end of the plotted range in figure \ref{fig:TemporalStability}(b), is almost the same as the $\omega_i(k=\pi)$ of the unconfined limit $q=\infty$ shown in figure \ref{fig:TemporalStability}(a). 
In the unconfined limit, the hyperbolic-cosecant terms (Eqs. \eqref{eq:defSeriesTerms}) diverge as $q^2/m^2$ when $q\rightarrow\infty$. 
Keeping these leading terms compensates the $q^2$ in the denominator of $A$ and $B$ (Eqs.\eqref{eq:defABinDispRel}) and 
one recovers the unconfined dispersion relation given by Saffman\cite{saffman1992}
\begin{equation}
 \omega_i=\pm\frac{k}{2}(2\pi-k), \label{eq:unconfinedDispRel}
\end{equation}
valid in the limit $q=\infty.$ 
Keeping extra terms give either vanishing or divergent contributions. 
Thus, we have not been able to obtain a correction of this dispersion relation in the case of weak confinement. 
This is probably related to how quickly the confined dispersion relation approaches that of the unconfied case.

\section{Spatio-temporal stability}\label{sec:spatiotemporal}

In practice, the absolute or convective (A/C) nature of an instability is determined in two different ways: finding the {\it absolute growth rate} from the complex dispersion relation \cite{huerrMonk1990} and from a numerically computed impulse response \cite{brancher1997,delbende1998}. 
Whereas the computation from the numerical impulse response involve simulations in sufficiently large domains and for sufficiently large times, it readily provides the perturbation growth rate in different reference frames. 
The determination of the A/C character of an instability in a given frame is generally more efficient from the dispersion relation, but computing the growth rate in different reference frames usually involves the recomputation of the dispersion relation for each particular frame.
This needs not to be the case, since the complex dispersion relation encodes the stability properties in all frames. 
Indeed, a change of reference frame in the dispersion relation corresponds to a simple Doppler shift. 
In the folowing, we present a simple but general geometric construction, akin to a Legendre transform, which allows to retrieve the growth rate of an instability in different reference frames from the dispersion relation computed on a single frame.

\subsection{Growth rate in different reference frames from the dispersion relation}\label{sec:LegendreGeoConstruction}

In the laboratory reference frame (at rest), the absolute growth rate is given by 
\begin{equation}
\sigma = \omega_i(k_0),
\label{eq:absGrowthRate}
\end{equation}
where $k_0$ is the absolute wavenumber that satisfies the zero group velocity condition
\begin{equation}
\frac{\mathrm{d} \omega}{\mathrm{d} k}(k_0) = 0.
\label{eq:SaddlePointConditionRestFrame}
\end{equation}
and the pinch point criterion~\cite{briggs1964,HuerreRossi1998}. 
If we consider now a reference frame moving at a velocity $v,$ the dispersion relation in the new frame is given by a Doppler shift
\begin{equation}
 \omega^v(k)=\omega(k) -v k, \label{eq:movingDispersionRelation}
\end{equation}
where $\omega$ is the the dispersion relation in the rest frame. 

In the moving frame, the absolute wavenumber $k^v_{0} = k^v_{0r} + i k^v_{0i}$ satisfying the zero group velocity condition is now given by 
\begin{equation}
\frac{\mathrm{d} \omega^v}{\mathrm{d} k}(k^v_{0})
=0, \label{eq:SaddlePointConditionMovingFrame}
\end{equation}
or equivalently from Eq. \eqref{eq:movingDispersionRelation}
\begin{equation}
  \frac{\mathrm{d} \omega}{\mathrm{d} k}(k^v_{0})=v,\label{eq:movinGroupVel}
\end{equation}
stating that the zero group velocity in the moving frame corresponds to a group velocity $v$ in the laboratory frame~\cite{HuerreRossi1998}. 
The growth rate 
 in the reference frame moving with velocity $v$ is then given as in Eq.\,\eqref{eq:absGrowthRate} by 
\begin{equation}
 \sigma(v)=\omega_i^v(k^v_0)=\omega_i(k^v_0)- v k^v_{0i}. \label{eq:growthRateMovingAtV}
\end{equation}
Thus, $\sigma(v)$ is usually obtained by calculating the Doppler-shifted dispersion relation $\omega^v$ separately for given values of $v$, then using \eqref{eq:SaddlePointConditionMovingFrame} to obtain $k^v_0,$ and finally the left equality in Eq.\,\eqref{eq:growthRateMovingAtV} to find the growth rates $\sigma(v)$ relating to these values of $v$. 

We now present an alternative method for obtaining the growth rate $\sigma$ directly {\it as a function} of $v.$ 
Assuming that $\omega(k)$ is an analytic function of $k,$ condition Eq.\,\eqref{eq:movinGroupVel} can be written in terms of the imaginary part of $\omega$ and its derivatives as
\begin{subequations}
\begin{align}
\frac{\partial \omega_i}{\partial k_i}(k^v_{0}) & = v, \label{eq:SaddlePointCondition1} \\
\frac{\partial \omega_i}{\partial k_r}(k^v_{0}) & = 0. \label{eq:SaddlePointCondition2}
\end{align}
\label{eq:SaddlePointCondition}
\end{subequations}
\begin{figure}
\centering
\begin{tabular}{ll}
\hspace{-.1cm}(a)\includegraphics[width=0.49\textwidth]{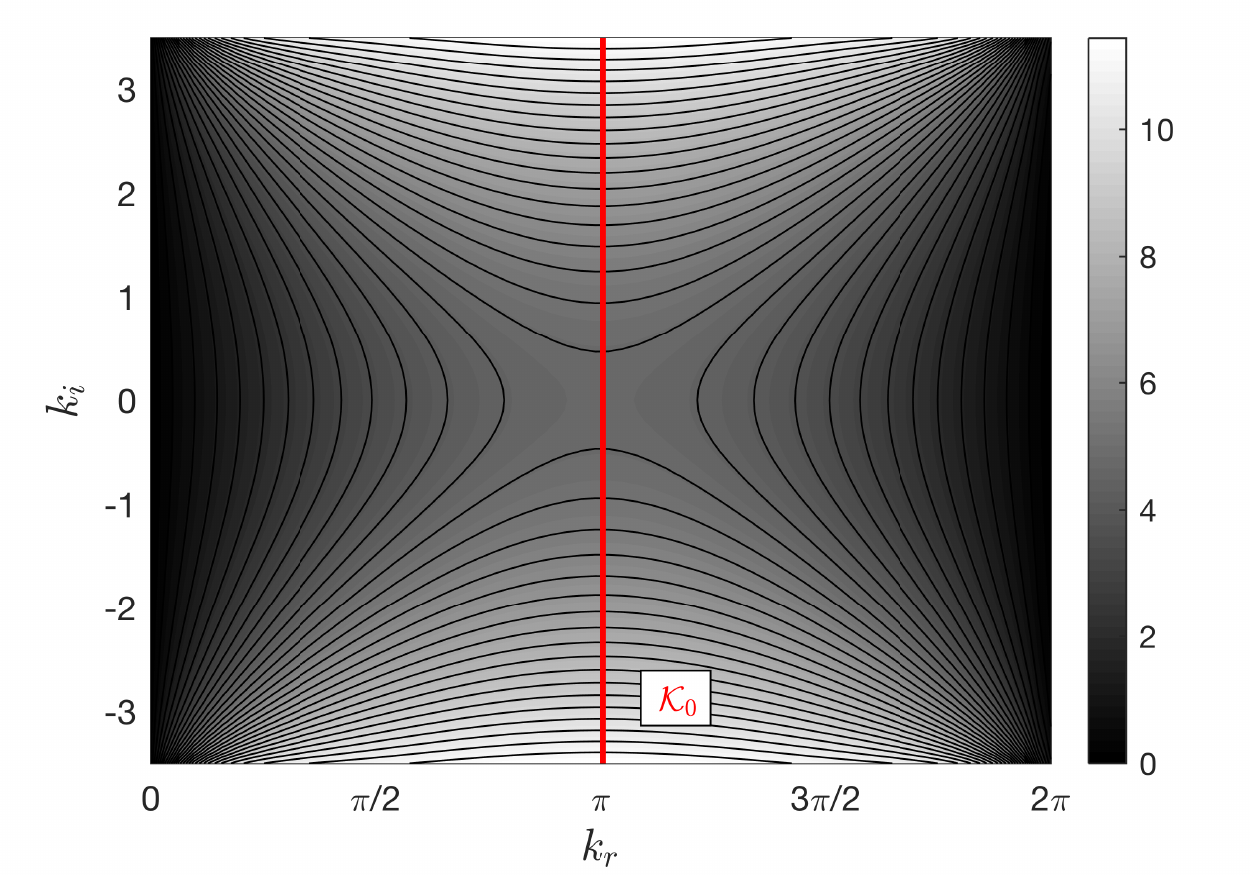} &
\hspace{-.1cm}(b)\includegraphics[width=0.49\textwidth]{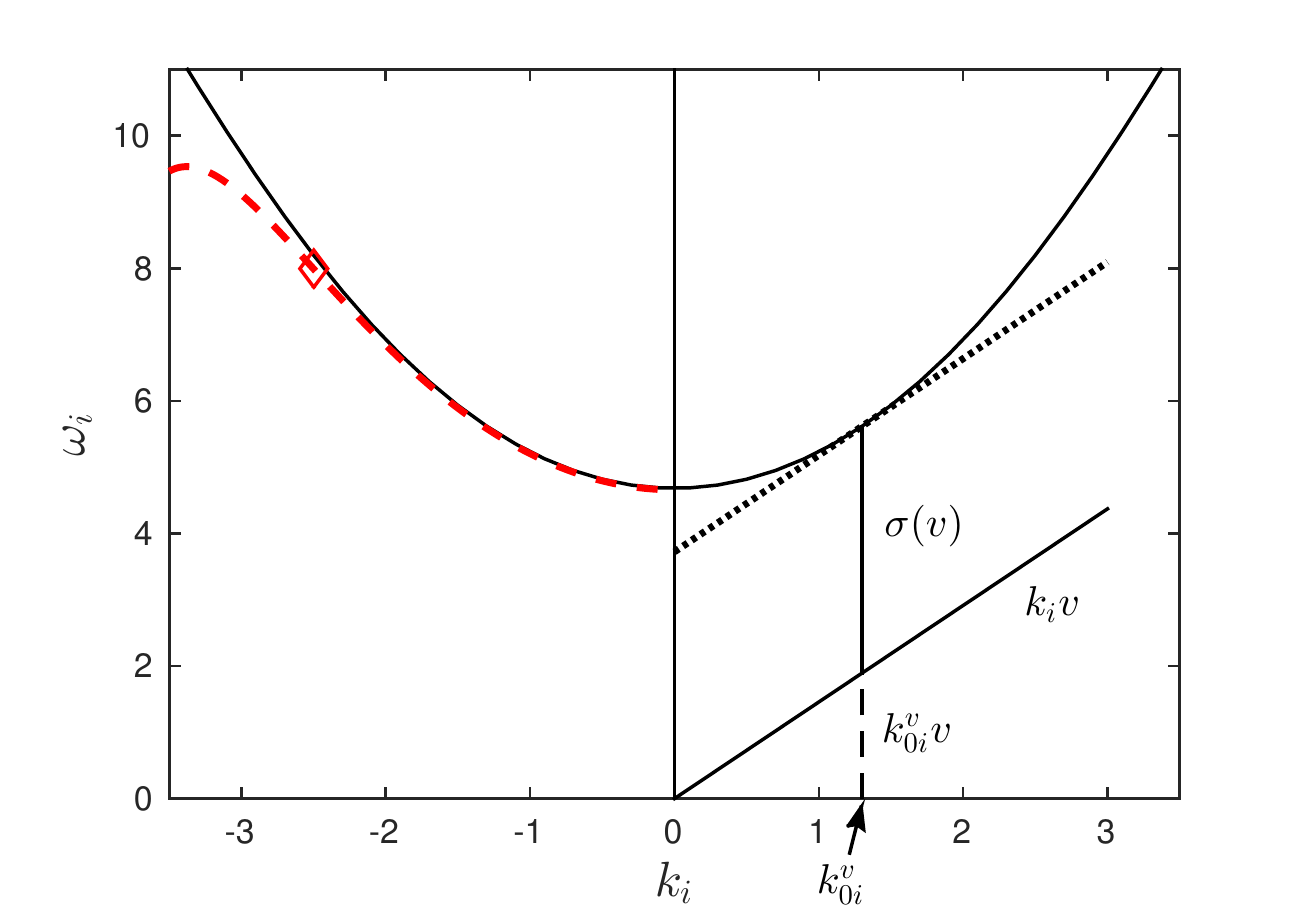}
\end{tabular}
\caption{(a) Isocontours of $\omega_i$ in the complex $k$-plane, calculated from the dispersion relation \eqref{eq:DispersionRelation} for $q = 0.8$. The solid red line represents the locus of absolute wavenumbers defined by \eqref{eq:LocusAbsoluteWavenumbers}.
(b) Geometrical construction to determine the growth rate $\sigma(v)$ from $\omega_i(k^v_{0r}(k_i),k_i)$. The $k^v_{0i}$ pertaining to a particular $v$ is given by \eqref{eq:SaddlePointCondition1}, and the corresponding growth rate is obtained from \eqref{eq:growthRateMovingAtV}. The dashed line on the left side of the plot shows the $\omega_i$ given by approximation \eqref{eq:limitqto0DispersionRelation}, with the diamond indicating the inflection point determining the validity limit of the approximation (see text).}
\label{fig:treatDispRelation}
\end{figure}
\noindent Imposing condition \eqref{eq:SaddlePointCondition2}, which is independent of $v,$ determines the locus of absolute wavenumbers 
\begin{equation}
\mathcal{K}_0 = \left\{ k^v_0 = k^v_{0r} + i k^v_{0i} : \frac{\mathrm{d} \omega}{\mathrm{d} k}(k^v_{0})=v \right\},
\label{eq:LocusAbsoluteWavenumbers}
\end{equation}
which is a (one dimensional) curve in the complex $k-$plane that contains the union of absolute wavenumbers in any moving frame. 

The locus $\mathcal{K}_0$ of absolute wavenumbers of the dispersion relation Eq.\,\eqref{eq:DispersionRelation} is shown as the solid red line in Figure \ref{fig:treatDispRelation}(a), together with the contour levels of $\omega_i(k)$ for $q=0.8.$ 
Due to the symmetry of the perturbations around $k=\pi$, condition \eqref{eq:SaddlePointCondition2} is satisfied for $k_r=\pi$ and $\mathcal{K}_0$ involves different values of $k_i$ with $k_r$ constant. 
It will be usually possible to parameterize $\mathcal{K}_0$ in terms of $k_i,$ although it can also happen that $\mathcal{K}_0$ contains multiple values of $k_r$ for some interval of $k_i,$ as shown in the example of Appendix \ref{ap:karmanLegendre}. 
In either case, the imaginary part of the frequency can be expressed from the dispersion relation as a (possibly multivalued) function of $k_i$ as $\omega_i(k^v_{0r}(k_i),k_i).$ 
Note that, when restricted to $k^v\in\mathcal{K}_0$, it follows from Eq.\,\eqref{eq:SaddlePointCondition2} that
\begin{equation}
 \mathrm{d} \omega_i = \frac{\partial \omega_i}{\partial k_i} \mathrm{d}k_i, \label{eq:variesOnlyWithKi}
\end{equation}
so that the imaginary frequency $\omega_i$ varies through the variation of $k_i$ only and not through $k^v_{0r}.$ 
Thus, it is natural to view $\omega_i(k^v_{0r},k^v_{0i})$ as a one variable function and $k_i$ as the independent variable. 
We can now re-write the growth rate in Eq.\,\eqref{eq:growthRateMovingAtV} as
\begin{equation}
 \sigma(v)=\omega_i(k^v_r(k_i),k_i) - vk_i, \label{eq:LegendreTransform}
\end{equation}
where $v=\frac{\partial \omega_i}{\partial k_i}.$ 
Eq.\,\eqref{eq:LegendreTransform} is in the form of a Legendre transform exchanging $\omega_i$ by the growth rate $\sigma$ and $k_i$ by its conjugate variable $v.$ 
As shown in Fig.~\ref{fig:treatDispRelation}(b), it corresponds to a simple geometrical construction to determine $k^v_{0i}$ and the growth rate $\sigma$ for the corresponding $v.$ 
The solid curve shows $\omega_i(k_r=\pi,k_i)$ from the dispersion relation \eqref{eq:DispersionRelation} for $q=0.8,$ and the straight line represents $k_i v$ for a given $v,$ i.e. the second term in Eq.\,\eqref{eq:LegendreTransform}. 
Condition \eqref{eq:SaddlePointCondition1} corresponds to choosing $k^v_{0i}$ as the $k_i$ for which the $\omega_i$ curve is parallel to this line. 
The growth rate along the spatiotemporal ray $x/t = v$ is then given by Eq.\,\eqref{eq:growthRateMovingAtV} as the vertical distance between $\omega_i$ and the straight line, evaluated at $k^v_{0i}.$

\begin{figure}
\centering
\begin{tabular}{ll}
\hspace{-.1cm}(a)\includegraphics[width=0.49\textwidth]{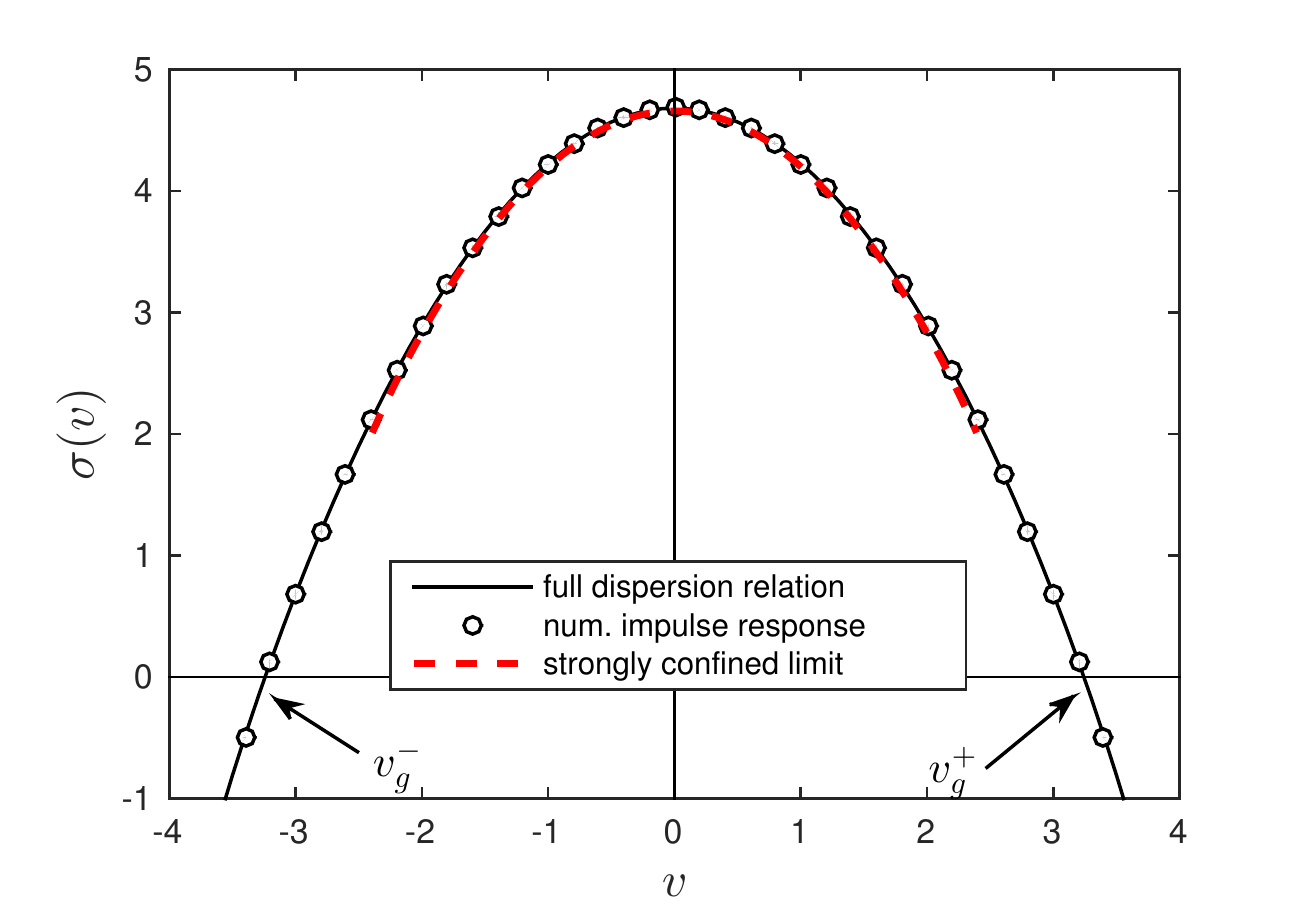} &
\hspace{-.1cm}(b)\includegraphics[width=0.49\textwidth]{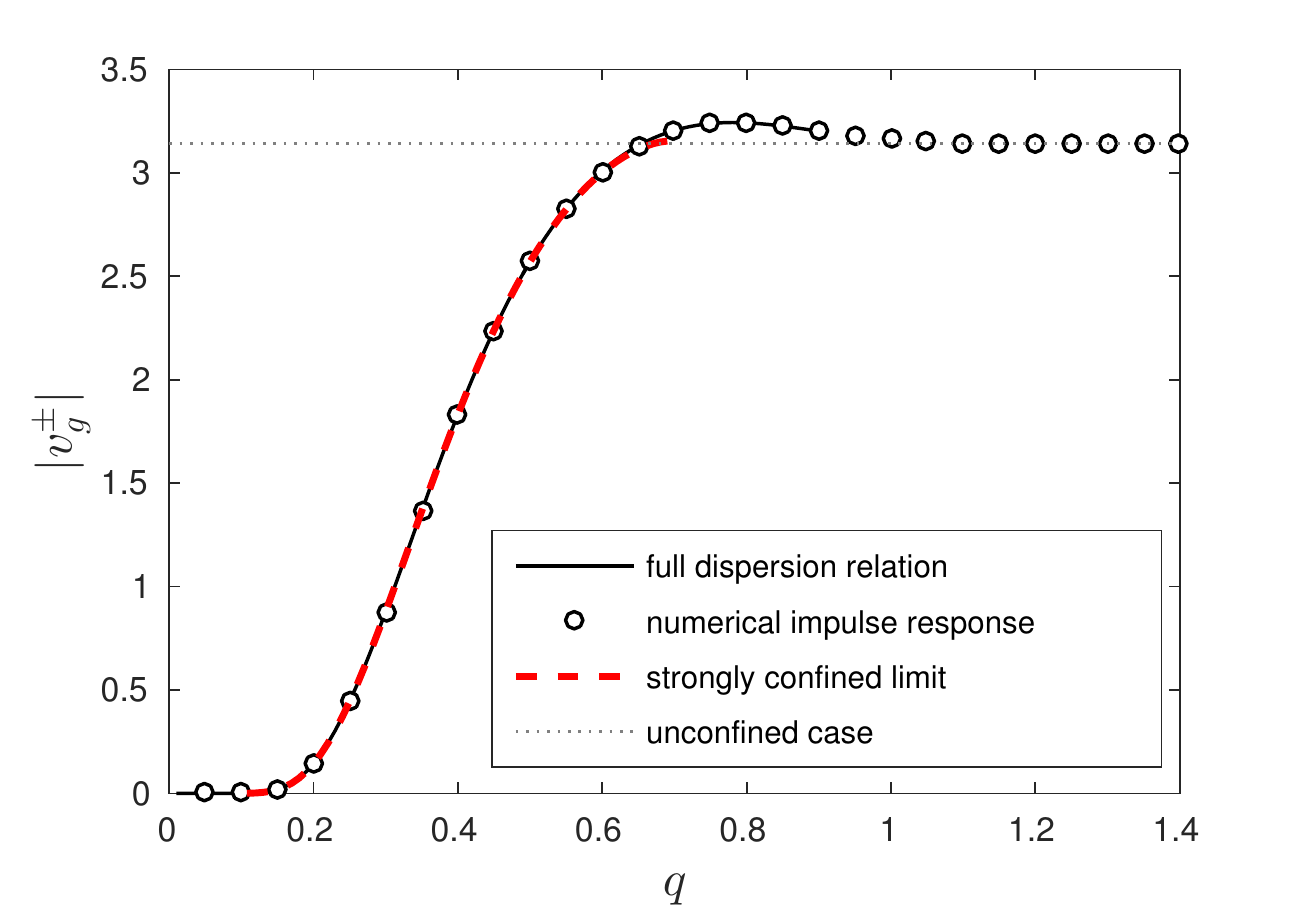}
\end{tabular}
\caption{(a) Growth rate $\sigma(v)$ of the impulse response wave packet along spatio-temporal rays $x/t = v$ for a confinement ratio $q = 0.8$. 
The streamwise extent of the wave packet is given by the leading- and trailing-edge velocities $v_g^\pm$ such that $\sigma(v_g^\pm) = 0$. Results from imposing the transformation \eqref{eq:LegendreTransform} to the exact (solid line) and aproximate (dashed line) dispersion relations and from numerical simulations of the impulse response (circles).
(b) Magnitude of the leading- and trailing-edge velocities $|v_g^\pm|$ of the impulse response wave packet versus the confinement ratio $q$. Results from the analytical dispersion relation with the graphical method (solid line) are compared with those from numerical simulations of the impulse response (circles). As $q$ increases, $|v_g^\pm|$ quickly approaches the theoretical value of $\pi$ for the unconfined single row of point vortices \cite{huerre1992}.}
\label{fig:instabilityPropagation}
\end{figure}

The solid line in Fig.~\ref{fig:instabilityPropagation}(a) shows $\sigma(v)$ resulting from the application of this technique to the dispersion relation ~\eqref{eq:DispersionRelation} for the case $q = 0.8$. 
Due to the upstream-downstream symmetry of the system, $\omega_i$ and $\sigma(v)$ are even respectively in $k_i$ and $v.$ 
Also shown in Fig.~\ref{fig:instabilityPropagation}(a), in dashed line, is the corresponding growth rate obtained from the dispersion relation\,\eqref{eq:limitqto0DispersionRelation} asymptotically valid in the limit of strong confinement. 
The agreement for the plotted values is very good. 
However, the approximate curve of $\sigma(v)$ does not extend to larger values of $|v|$ because the corresponding $\omega_i(k^v_0(k_i))$ presents an inflection point corresponding to a maximum of $|\frac{\partial \omega_i}{\partial k_i}|,$ shown with a diamond in the left side of Fig.~\ref{fig:treatDispRelation}(b). 
Beyond this point the approximation given by Eq.~\eqref{eq:limitqto0DispersionRelation} is meaningless. 

For large times, the spatio-temporal region of growth of an initial impulse localized at $(x,t)=(0,0)$ is given by the spatio-temporal rays $x/t$ of leading- and trailing-edge velocities $v_g^\pm=x/t$ such that $\sigma(v_g^\pm) = 0$. 
Note that applying this marginal stability condition $\sigma(v_g^\pm) = 0$ in Eq.~\eqref{eq:LegendreTransform} recovers Dee \& Langer\cite{DeeLangerPRL1983FrontVelocitySelection} and van Saarlos\cite{vanSaarlos87} condition $v_g=\mathrm{d}\omega/\mathrm{d}k=\omega_i/k_i$ for the linearly selected velocity of front propagation. 

The magnitude of the leading- and trailing-edge velocities $|v_g^\pm|$ is reported in Fig.~\ref{fig:instabilityPropagation}(b) for a range of confinement ratios. 
The solid and dashed lines show the values obtained from the full dispersion relation~\eqref{eq:DispersionRelation} and from its strong confinement approximation~\eqref{eq:limitqto0DispersionRelation}, respectively. 
In the strong confinement limit this propagation velocity decreases rapidly when $q\lesssim 0.6,$ trend which is quantitatively well captured by the approximated dispersion relation~\eqref{eq:limitqto0DispersionRelation} (dashed line). 
This approximation works well until $q=0.69,$ after which the approximation looses validity before the the marginal stability criterion is satisfied. 
For sligthly larger confinement ratios the full dispersion relation~\eqref{eq:DispersionRelation} (solid line) reveals that $|v_g^\pm|$ reaches a maximum of $|v_g^\pm|=3.243$ around $q = 0.78.$ 
Remarkably, this value is larger than that of the unconfined case~\cite{huerre1992}, which is given by $|v_g^\pm|=\pi$ and represented by the horizontal dotted line in Fig.~\ref{fig:instabilityPropagation}(b). 
This increase in the propagation velocity of the instability is a destabilizing effect of confinement analogous to that found for parallel flows, as we will discuss in detail below in \ref{sec:tanhComparison}. 
Past this maximum in $|v_g^\pm|$, a small decrease can be observed before the evaluation fails because of the divergence of the series in the dispersion relation~\eqref{eq:DispersionRelation}. 
Indeed, the terms of the series grow exponentially with $m$ when $|k_i|>\pi/q,$ rendering the evaluation imposible through a direct summation for values of $q$ greater than $0.9$. 
Therefore, in order to complete the curve we turn in the next section to a different approach that relies on direct numerical simulations of the impulse response.

\subsection{Growth rates from the numerically computed impulse response}

The asymptotic properties of the impulse response wave packet can also be retrieved from a numerical simulation of the response of the system to a localized initial perturbation \cite{brancher1997,delbende1998}. 
This method was recently implemented for the secondary instability of the confined K\'arm\'an street \cite{mowlavi2016}, using a point vortex model closely related to the system presently under study. 
Hence we adopt a similar approach and compute the time evolution of the single street of vortices through direct time integration of the linearized perturbation equations \eqref{eq:LinearizedEquationsMotion}. 
We use a finite number of physical vortices and simulate the infinite series in the perturbation equations through additional virtual vortices that are slaved to the physical ones (for additional details see \cite{mowlavi2016}). 
The equations of motion \eqref{eq:LinearizedEquationsMotion} are then applied to the physical vortices, taking into account the velocities induced by the virtual vortices, and are advanced in time with an explicit Euler scheme. 
This procedure is implemented in MATLAB with a nondimensional time step $\Delta t = 0.1$, $M = 201$ physical vortices and $6M$ virtual vortices. 
The simulation is initialized with a small vertical displacement of the center physical vortex and the total integration time is $20$. 
The localized initial perturbation generates a growing wave packet whose amplitude is defined as
\begin{equation}
A(x,t) = \sqrt{x_m(t)^2+y_m(t)^2},
\end{equation}
where $m = \mathrm{round}(x)$. 
The growth rate observed along spatio-temporal rays $x/t = v_g$ emerging from the initial location of the perturbation can be evaluated from the amplitude at two distinct time instants $t_1$ and $t_2$ via
\begin{equation}
\sigma(v_g) = \frac{1}{t_2-t_1} \ln \left[ \frac{A(v_g t_2,t_2) \sqrt{t_2}}{A(v_g t_1,t_1) \sqrt{t_1}} \right].
\end{equation}

A simulation with confinement ratio $q = 0.8$ is first performed and the growth rate of the resulting wave packet is shown in circles in Fig.~\ref{fig:instabilityPropagation}(a). 
The excellent agreement obtained between the numerical growth rate and that from the analytical dispersion relation validates the accuracy of the numerical method. 
We now carry out simulations of the impulse response for a range of confinement ratios and retrieve the leading- and trailing-edge velocity magnitude $|v_g^\pm|$ in each case. 
The results are displayed in circles in Fig.~\ref{fig:instabilityPropagation}(b) and again compare extremely well with the velocities obtained previously from the analytical dispersion relation. 
The range for $q$ is no longer limited at $0.9$ and we can compute how the curve of $v_g^\pm$ quickly approaches the theoretical value of $\pi$ deduced by Huerre\cite{huerre1992} and previously reported in Ref.~\cite{brancher1997} for the unconfined single row of point vortices. 

\section{Significance for free shear layers}\label{sec:applications}
As mentioned in the introduction, a row of vortices results from the saturation of the primary instability of a free shear layer~\cite{DrazinReid2004}. 
In this context, the instability of the point vortex model studied in the previous section can be regarded as the secondary instability of the mixing layer. 
However, since primary and secondary instabilities do not necessarily have the same absolute/convective character, they may be affected differently by confinement; this will have implications for the development of the shear layer. 
In the present section we address what are these possible implications. 
We begin with a comparison of the spatio-temporal properties of the aforementioned instabilities. 

\subsection{Comparison with the primary instability of a $\tanh$ profile} \label{sec:tanhComparison}

\begin{figure}
\includegraphics[width=0.5\textwidth]{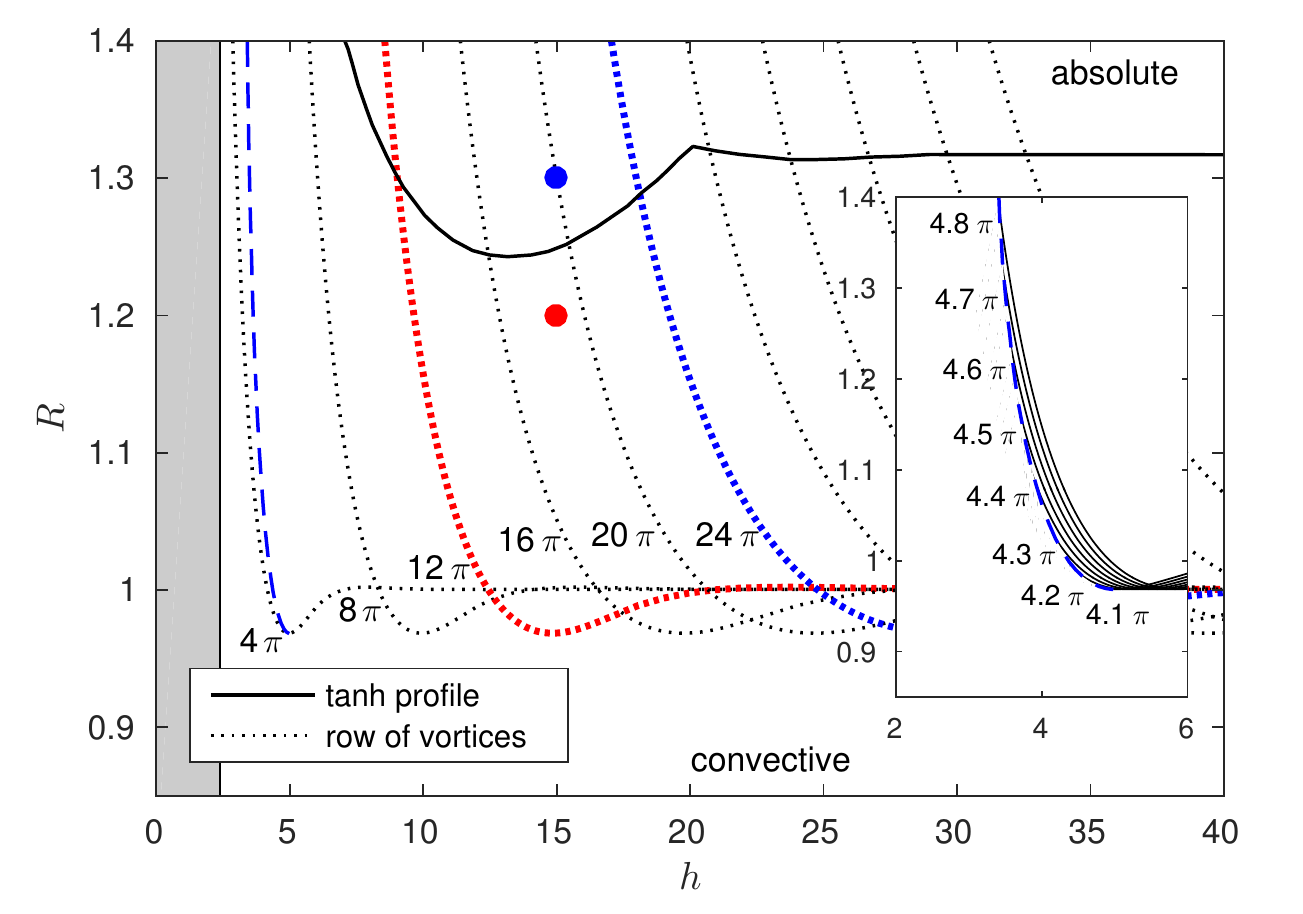}
\caption{Spatio-temporal behavior of the $\tanh$ profile \eqref{eq:MixingLayerProfile} and the row of point vortices. 
Domains of absolute and convective primary instability ($\tanh$ profile) are separated by the critical velocity ratio $R_{c1}$ (solid line), while in the grey region it is stable. 
Domains of absolute and convective instability of the row of vortices, separated by the critical velocity ratio $R_{c2,\lambda},$ are shown for different values of the inter-vortex spacing $\lambda = 4\pi,8\pi,12\pi,16\pi,\ldots,44\pi$ (dotted lines). 
In both cases, values of $R$ greater (lower) than the reported critical value lead to absolute (convective) instability. 
The $R_{c2,\lambda}$ determine the critical velocity ratio of the secondary instability to the right of $\bar{R}_{2C}$ (dashed line), below which the secondary instability is always convective. 
The inset shows a close-up around $\bar{R}_{2C}$ including various intersecting $R_{c2,\lambda}$ (solid lines).  
Results for the primary instability from Healey \cite{healey2009}.
}
\label{fig:ConvectiveAbsoluteBehavior}
\end{figure}

Healey \cite{healey2009} considered the inviscid instability of a confined plane mixing layer profile of the form
\begin{equation}
U(y) = 1 + R \tanh \frac{y}{2},
\label{eq:MixingLayerProfile}
\end{equation}
where $U$ is the base flow velocity in the $x$ direction, $y$ is the cross-stream direction and $R = (U_1^*-U_2^*)/(U_1^*+U_2^*)$ is a velocity ratio with $U_1^*$ and $U_2^*$ the dimensional velocities far above and below the mixing layer. 
Velocities are nondimensionalized with the average advection velocity $\bar{U}^* = (U_1^*+U_2^*)/2$, and lengths are nondimensionalized with the shear layer thickness $\delta$. 
Symmetric confinement by horizontal parallel plates is enforced by free slip boundary conditions at $y = \pm h.$ 
Note that $2h\delta=d.$ 
In this setting, Healey showed that confinement has a stabilizing effect on the temporal stability of the mixing layer, but can increase the region of absolute instability in a certain range of $h$. 
The critical velocity ratio $R_{c1}$ ($c$ and $1$ for `critical' and `primary') that separates regions of absolute and convective instability obtained by Healey for the tanh profile Eq.~\eqref{eq:MixingLayerProfile} is shown by the solid line in the $(h,R)$ plane in Fig.~\ref{fig:ConvectiveAbsoluteBehavior}. 
For the strong confinement shown in the shaded region of Fig.~\ref{fig:ConvectiveAbsoluteBehavior} the primary instability is no longer present. 

Returning to our model of point vortices, the increase of $|v_g^\pm|$ with increasing confinement observed in Fig.~\ref{fig:instabilityPropagation}(b) indicates a similar destabilizing effect of confinement on the secondary instability of the mixing layer. 
In order to compare quantitatively the convective/absolute behavior of this secondary instability with the primary instability results from Fig.~\ref{fig:ConvectiveAbsoluteBehavior}, we need to relate dimensional quantities in our point vortex model with their mixing layer counterparts. 
If the single row of vortices emerges out of the saturating mixing layer, we can equate the circulation per unit length of both systems to get
\begin{equation}
\frac{\Gamma}{a} = - (U_1^*-U_2^*).
\label{eq:DimensionalQuantitiesRelationship}
\end{equation}
Placing ourselves in the laboratory frame, in which the vortices are advected at the mean velocity of the mixing layer $\bar{U}^*$, the dimensional front velocities of an impulse response wave packet in the row of vortices is
\begin{equation}
v_f^{\pm*} = \bar{U}^* + v_g^{\pm*} = \bar{U}^* - \frac{\Gamma}{2\pi a}  v_g^{\pm},
\label{eq:DimensionalFrontVelocities}
\end{equation}
where $v_{f,g}^{-}<v_{f,g}^{+},$ the $v_{g}^{\pm*}$ are the dimensional leading- and trailing-edge velocities of the impulse wave packet in the reference frame of the vortices, and $v_g^{\pm}$ are their nondimensional counterparts displayed in Fig.~\ref{fig:instabilityPropagation}(b). 
We now nondimensionalize \eqref{eq:DimensionalFrontVelocities} with $\bar{U}^*$ and make use of \eqref{eq:DimensionalQuantitiesRelationship} to obtain
\begin{equation}
v_f^{\pm} = 1 + \frac{U_1^*-U_2^*}{2\pi \bar{U}^*} v_g^{\pm} = 1 + \frac{R}{\pi} v_g^{\pm}.
\end{equation}
The row of vortices becomes convectively unstable in the laboratory frame when $v_f^-$ turns positive. 
This allows us to formulate a critical velocity ratio for the secondary pairing instability $R_{c2},$ above which the single row of point vortices undergoes convective to absolute transition
\begin{equation}
R_{c2} = \frac{\pi}{|v_g^-|}.
\label{eq:CriticalVelocityRatio}
\end{equation}
Finally, in order to compare $R_{c1}$ with $R_{c2},$ we need to relate lengths from the point vortex model to those in the mixing layer. 
Physically, it appears at first reasonable to equate the dimensional inter-vortex spacing $a$ with the wavelength of the temporally most unstable primary perturbation of the mixing layer. 
However, this would amount to prescribing a unique wavelength selection to the primary instability, whereas in real mixing layers the wavelength selection is a complex dynamical process involving various factors. 
For now, we leave $a$ as an arbitrary parameter, and we consider first a general row of vortices independently of how it was formed. 

We relate the aspect ratio $q$ of our model to the length scale of the mixing layer through the distance between the confining plates at $y=\pm h$ as 
\begin{equation}
q = \frac{d}{a} = \frac{2\delta h}{a} = \frac{2h}{\lambda},
\label{eq:AspectRatio}
\end{equation}
where the inter-vortex spacing is now given by $\lambda = a/\delta$. 
For any given vortex separation distance $\lambda$ one can combine \eqref{eq:CriticalVelocityRatio} and \eqref{eq:AspectRatio} with the data from Fig.~\ref{fig:instabilityPropagation}(b), which determines a family of critical velocity ratios $R_{c2,\lambda}(h)$ that separate regions of absolute and convective vortex pairing instability in the $(h,R)$ plane. 
Critical curves $R_{c2,\lambda}$ for different values of the inter-vortex spacing $\lambda = 4\pi,8\pi,12\pi,16\pi,\ldots,44\pi$ are plotted in dotted lines in Fig.~\ref{fig:ConvectiveAbsoluteBehavior}. 
A remarkable aspect of these $R_{c2,\lambda}$ curves is the possibility of absolute instability with $R<1,$ that is, without counterflow. 
This promoted absolute intability takes place around $h\approx\lambda/2.5.$

According to the results plotted in Fig.~\ref{fig:ConvectiveAbsoluteBehavior}, all possible combinations between absolute or convective instability for the tanh profile and pairing instabilities are in principle possible for different values of $\lambda, R$ and $h.$
For example, if $(h,R)=(15,1.3)$ and $\lambda=24\pi,$ depicted in blue, the primary instability of the tanh profile is absolute while the (would-be secondary) vortex pairing instability of the row of vortices is convective.
We refer to this situation as 1A2C.
If we then change the $R$ and $\lambda$ values to $R=1.2$ and $\lambda=12\pi,$ depicted in red, we would have the opposite situation in which the tanh profile instability is convective and the instability of the row of vortices is absolute (1C2A).
The remaining possibilities 1A2A (both instabilities absolute) and 1C2C (both convective) are always present in the $(h,R)$-plane. 

Let us now consider the possible sequence of instabilities in a mixing layer, i.e., we now view the pairing instability of the vortex row as a secondary instability forming from the outcome of the primary instability of the shear layer.
Given the appropriate initial conditions or forcing, it is in principle conceivable that a vortex row be formed with any inter-vortex spacing lying within the range of unstable wavenumbers of the primary instability. 
Thus, the relevant range of $\lambda$ can be taken as given by the range of temporally unstable wavenumbers of the $\mathrm{tanh}$ profile. 
In the confined case this unstable range is given by Healey~\cite{healey2009} (see his Fig.~1) and goes from $k=2\pi/\lambda=0$ up to an $h-$dependent critical wavenumber $k_c(h)$ which decreases with $h$ until it goes to zero at $h\approx 2.399.$ 
This yields the stable shaded region in Fig.~\ref{fig:ConvectiveAbsoluteBehavior}. 
In Fig.~\ref{fig:ConvectiveAbsoluteBehavior}, the plotted $R_{c2,\lambda}$ curves correspond to wavenumber values $k=1/2,1/4,1/6,\ldots,1/22$.
Apart from $k=1/2,$ which is the critical wavenumber $k_c$ in the unconfined case, these values are within the range of temporally unstable wavenumbers when $h$ is not too small. 
Thus, their corresponding $R_{c2,\lambda}$ curves are potentially valid critical curves for the secondary instability of a $\mathrm{tanh}$ shear layer, unlike the $\lambda=4\pi$ curve since the wavelength issuing from the primary instability must be larger. 
The region of validity for the secondary instability of the $R_{c2,\lambda}$ family of critical curves of the vortex row is limited by the dashed line of Fig.~\ref{fig:ConvectiveAbsoluteBehavior}, which translates to the present context the critical wavelength $k_c(h)$ of the $\mathrm{tanh}$ profile given by Healey~\cite{healey2009}. 
We denote the corresponding function as $\bar{R}_{2C}(h)$ since for $R$ below its curve the secondary instability is necessarily convective, as explained in detail in Appendix \ref{ap:ExplainR2C}. 

\begin{figure}
\includegraphics[width=0.5\textwidth]{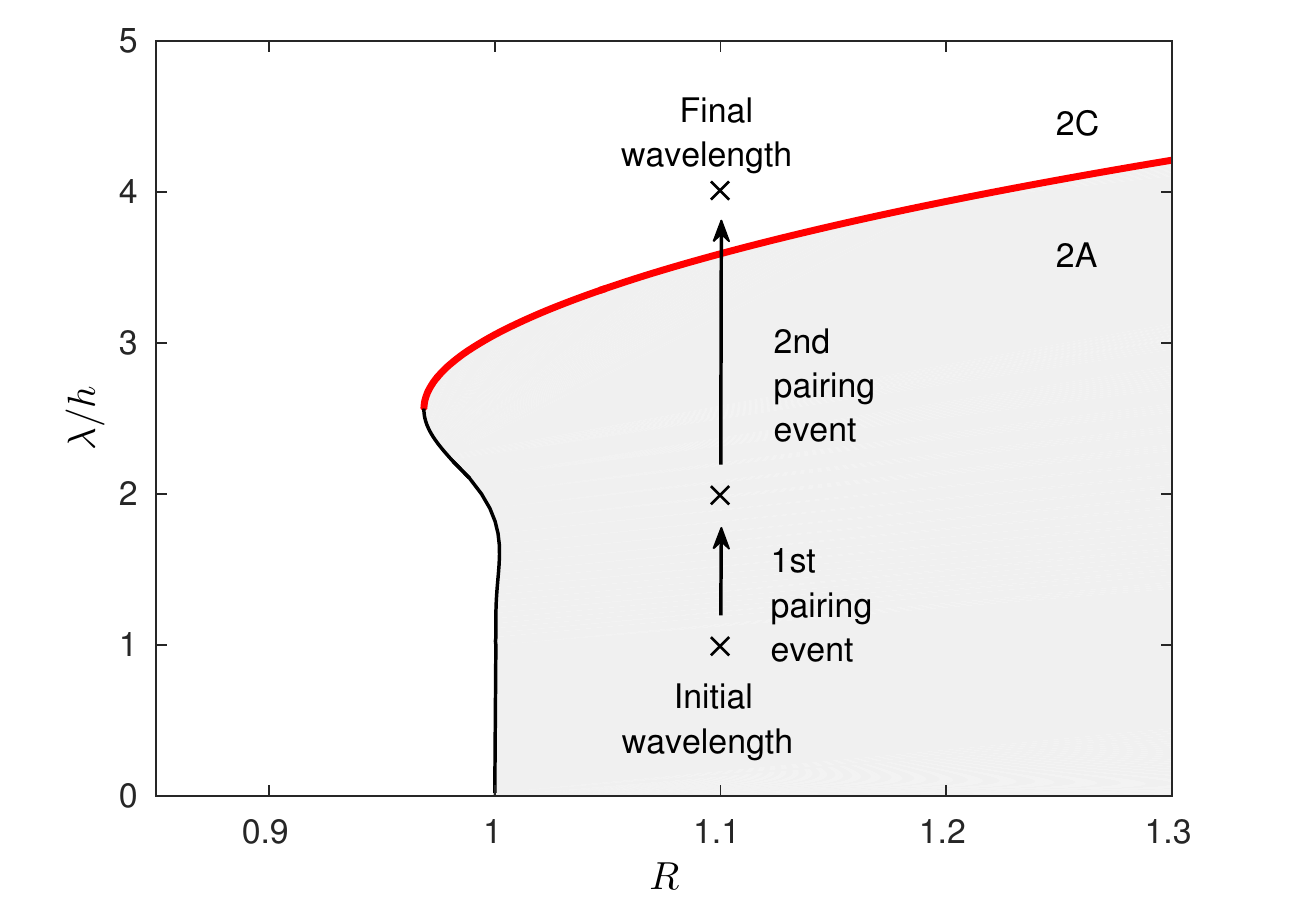}
\caption{Tentative scenario for wavelength selection. If the ratio $\lambda/h$ of inter-vortex spacing to mid-channel width is such that the secondary instability is absolute (2A), successive vortex pairings will increase $\lambda$ and eventually bring the system in the region where the secondary instability is convective (2C), in which it will remain. According to this interpretation, a spatially developing mixing layer with counterflow $(R>1)$ can only exhibit a row of vortices if $\lambda/h$ is large enough to fall in the region 2C. }
\label{fig:waveSelection}
\end{figure}

\subsection{Discussion} 
Above and to the right of $\bar{R}_{2C}$ it is, depending on $\lambda,$ still possible to have all combinations between absolute and convective for the primary and secondary instabilities.
Among these cases, the most commonly observed in co-flowing mixing layer experiments is 1C2C, for which the primary and a number of secondary instabilities succeed each other spatially in a facility dependent way due to a high sensitivity to incoming noise\cite{hoHuerr1984}.
In this case, vortex merging events will increase the wavelength $\lambda$ thus selecting curves $R_{c2,\lambda}$ which move successively to the right in Fig.~\ref{fig:ConvectiveAbsoluteBehavior} and the instability generally remains convective. 

The same situation for the secondary instability would occur in case 1A2C, except that the initial wavelength $\lambda_0$ of the vortex row would be robustly selected by the absolute primary instability. 
This selected wavelength $\lambda_0$ can be expected to correspond to the real part of the absolute wavenumber of the primary instability $k_0,$ although non-linear effects, strong forcing or initial conditions could also have an effect. 
Thus, it would be possible to define a unique critical curve $R_{c2,\lambda_0}$ that would be valid in the region 1A. 
However, this involves finding in the complex $k$-plane the absolute wavenumber of the $\tanh$ profile as a function of $R$ and $h,$ which goes beyond the scope of the present paper.
We proceed instead to discuss the dynamical consequences that can be expected from considering a varying $\lambda.$

As we mentioned above, vortex merging events will increase the wavelength $\lambda$ while moving the relevant critical curve $R_{c2,\lambda}$ to the right.
This can be better represented in the $(R,\lambda/h)$-plane of Fig.~\ref{fig:waveSelection}, in which the family $R_{c2,\lambda}$ separating the regions of absolute and convective instability for the row of vortices collapses into a single critical curve. 
Increasing $\lambda$ due to vortex merging translates then directly into moving upwards in the $(R,\lambda/h)$-plane. 
Fig.~\ref{fig:waveSelection} shows that every point in the region 2A is located below a region 2C. 
Thus, if an initial $\lambda/h$ is such that the secondary instability is absolute (2A), increasing $\lambda$ through successive vortex pairings will eventually bring the system into the convective region 2C, in which it will remain for subsequent mergers. 
An example of this process is depicted in Fig.~\ref{fig:waveSelection}, in which the initial $\lambda/h$ is doubled twice while undergoing two pairings before getting out of the region 2A into the 2C.

The situation depicted in Fig.~\ref{fig:waveSelection} is, however, not so simple for spatial shear layers; it falls in the not so straightforward situation of a secondary absolute instability.
The first thing to note is that the process of finite successive pairings drawn in Fig.~\ref{fig:waveSelection} is a transient one, so it should be observed as depicted if the `initial wavelength' within the 2A region results, for example, from a sudden modification of the confinement or other properly tailored conditions.
But the question remains about the consequences of an absolute secondary instability for a proper spatial shear-layer, that is, one in a statistically steady state.

One possibility of this absolute secondary instability would be the scenario of Chomaz\cite{Chomaz2004EJMBSecondaryAbsoluteAA,ChomazARFM2005}, discussed in the introduction, of a sudden one step transition to a complex behaviour.
Examples discussed in Ref.~\cite{Chomaz2004EJMBSecondaryAbsoluteAA} include the emergence of low frequency oscillations of the saturated nonlinear state of a complex Ginzburg-Landau model (on this model see also Ref.~\cite{CouaironChomaz1ary2dary1999}) and a mixing layer. 
These are reminiscent of low frequency modulations observed in 2D wakes\cite{biancofiore_gallaire_pasquetti_2011} or forced capillary jets \cite[\S IV.B]{horvathArrCorPoF2015}. 
It is plausible that the periodic or random appearance of these low frequency modulations could be related to the absolute or convective nature of the primary instability. 
That is, a 1A2A scenario with a subharmonic secondary instability could lead to periodic low frequency modulations of the saturated pattern of the primary instability, while the low frequencies in a 1C2A would be sensitive to inlet disturbances and more likely to appear random. 
A detailed assessment of these considerations requires dedicated studies of each particular case.

A dedicated study of the present case would have to involve nonlinear simulations or experiments, going beyond the scope of the present paper. 
However, in view of our present results, we can reassess the mixing layer results presented by Chomaz~\cite{Chomaz2004EJMBSecondaryAbsoluteAA} in support of the scenario of a one step transition to complex behavior. 
In his Fig.~8, Chomaz shows successive images of a shear layer with $R=1.42$ at Reynolds number $400$ kept artificially parallel by a diffusion-cancelling body force, and confined with freely slipping walls. 
The initially parallel shear layer is seen to develop KH vortices, first at its downstream end and subsequently moving upstream. 
The details of the upstream displacement of the emerging row of vortices cannot be ascertained from the images shown, but vortex pairings are clearly present and the distance between consecutive vortices increases in time. 
About the resulting flow, Chomaz notes that ``[t]he Global Mode for the mixing layer does not seem to be stable''~\cite{Chomaz2004EJMBSecondaryAbsoluteAA}, and he describes it as displaying random pairings (in Refs.~\cite{Chomaz2004EJMBSecondaryAbsoluteAA,ChomazARFM2005}, see the last quotation of the introduction). 
Based on the results for the absolute/convective instability of Stuart vortices without confinement~\cite{brancher1997}, Chomaz attributes this randomness to the alleged absolute nature of the pairing instability, analogous to what was observed on the Ginzburg-Landau model~\cite{CouaironChomaz1ary2dary1999,Chomaz2004EJMBSecondaryAbsoluteAA}. 

We can now provide a different interpretation of these results. 
Indeed, it is likely that this transient process shown in Fig.~8 of Ref.~\cite{Chomaz2004EJMBSecondaryAbsoluteAA} corresponds to the proposed process depicted in Fig.~\ref{fig:waveSelection}. 
Chomaz' numerical experiment with an initially parallel shear layer with counterflow seems precisely like the kind of properly tailored conditions in which one would expect to observe the occurence of this process. 
The parallel shear layer is first subject to the primary instability which, by the snapshot at time $300$ of Fig.~8 of Ref.~\cite{Chomaz2004EJMBSecondaryAbsoluteAA}, has developed into an absolutely unstable row of vortices that is subsequently subject to pairing through an absolute secondary instability. 
Vortex separation increases after pairing, and one can calculate the confinement ratio from the final snapshot at time $500$ of the figure. 
The separation $\lambda$ between the closest pair of vortices is then about $4$ times the mid-width $h$ of the channel. 
Considering the velocity ratio $R=1.42$ and the finite size of the vortices that favors convective instability with respect to point vortices (Fig.~4 of Ref.~\cite{brancher1997} or Fig.~12 in~\cite{Chomaz2004EJMBSecondaryAbsoluteAA}), this value is very likely close to the absolute/convective instability threshold of a more realistic row of vortices, see Fig.~\ref{fig:waveSelection}. 
Thus, it is possible that the final state of the spatial shear layer observed by Chomaz~\cite{Chomaz2004EJMBSecondaryAbsoluteAA} is only convectively unstable (case 1A2C), and that the observed randomness is a result of the high sensitivity to small disturbances (as in the case of the `secondary vortex street' studied in Ref.~\cite{kumar_mittal_2012}). 
This exemplifies what is, in our view, the most likely consequence of the present results.

\subsubsection*{A wavelength selection mechanism}
This possible outcome is consistent with the fact that the saturation of the instability of the row of vortices leads to a similar row with double the wavelength, which falls also in the unstable range of the primary instability. 
Thus, it could be possible for the primary instability to by-pass the wavelengths that yield absolute secondary instability by spatially saturating directly on a wavelength which is large enough so that the secondary instability is convective. 
In this situation, the practical consequence of the region of absolute secondary instability is to restrict the range of possible wavelengths present in the permanent state of a shear layer with counterflow. 
In other words, a row of vortices can appear on a spatial shear layer if it is convectively unstable, but not if it is absolutely unstable. 
According to this wavelength selection mechanism, a spatially developing mixing layer can only exhibit a row of vortices with wavelength $\lambda$ if $\lambda/h=2/q$ is large enough to fall in the region 2C. 
Therefore, we would say that vortex rows in spatial mixing layers with counterflow ($R>1$) can only occur below a small enough confinement ratio $q=2h/\lambda$ given by twice the inverse of the critical line in Fig.~\ref{fig:waveSelection}, i.e., when the confinement is strong enough. 

One consequence of this wavelength selection mechanism is the impossibility of having a spatial row of vortices with zero mean flow, i.e., in the limit $R=\infty$. 
This is related to the fact that the temporal instability does not disappear with confinement, but this situation could change in a more realistic viscous model.  
More interestingly, another consequence is the impossibility of creating spatial mixing layers with finite counterflow ($R>1$) in the absence of confinement.
Indeed, according to the critical curve of Fig.~\ref{fig:waveSelection}, for $1<R<1.3$ the distance between consecutive vortices should be more than about $3$ to $4$ times the distance from the center of the shear layer to the symmetrically confining walls. 
This would explain the failure of the attempt to obtain a spatial mixing layer by Humphrey \& Li\cite{humphreyLi1981tilting}.
While they report observations with the same flow rate in both directions, i.e., with zero mean flow, they also note that \textquotedblleft [u]nequal flow rate observations did not differ fundamentally from the results presented.\textquotedblright\,
The reported streamwise length of their free shear layer was of $7.62\,\mbox{cm},$ only slightly larger than the distance of $5.08\,\mbox{cm}$ to the confining walls.
This is not enough to accomodate even a pair of vortices that would not be absolutely unstable. 
Due to the different geometry, these results do not apply to the experiments of Strykowski and collaborators~\cite{StrykowskiNiccum1991,ForlitiEtal2005}: the celebrated experiment of Ref.~\cite{StrykowskiNiccum1991} was on cylindrical jets, and subsequent experiments on plane mixing layers with counterflow use asymmetric and non parallel confining walls~\cite{ForlitiEtal2005}. 
Interestingly, our results do suggest that it could be possible to experimentally generate plane shear layers with counterflow, only that for this to be achieved the mixing layer needs to be sufficiently long in the streamwise direction with respect to the vertical distance between the confining walls. 
The possibly disrupting effect of boundary layers at the confining walls can be minimized with moving walls. 
Recent experiments~\cite{BonifaceEtalEPL2017} showing the stabilizing effect of confinement on vortex streets could be adapted to test this prediction. 

\section{Summary and conclusions}\label{sec:conclusion}

We have characterized the effect of confinement on the absolute/convective nature of the pairing instability based on the infinite row of point vortices. 
This is a minimal model to study how the growth of free shear layers through successive pairings can be eventually limited by the effects of confinement. 
Surprisingly, as shown in Sec.~\ref{sec:TemporalStability}, the pairing instability is never fully stabilized by confinement in this conservative model, the temporal growth rate of the instability becoming instead exponentially small as confinement or, equivalently, vortex separation increases. 
This behaviour is well captured with an approximate expression for the dispersion relation (Eq.~\eqref{eq:limitqto0DispersionRelation}) that is asymptotically valid in this limit of strong confinement $q\rightarrow 0.$ 

This asymptotically valid dispersion relation also quantitatively captures the critical velocity marking the threshold between absolute and convective instability for values of confinement up to $q\approx0.69.$ 
For computing the thresholds, the growth rate of the impulse response as observed in different reference frames is obtained from the (full and approximate) dispersion relations in a single frame. 
This is done with a novel approach, described in Sec.~\ref{sec:LegendreGeoConstruction}, whereby the imaginary part of the wavenumber $k_i,$ and the relative velocity of a moving frame $v,$ are related as conjugate variables through a Legendre transformation of the growth rate $\omega_i$ evaluated on the locus of absolute wavenumbers. 
The obtained results are confirmed with numerical simulations of the linearized impulse response. 
These simulations do also extend the computed results to the moderate and weak confinement regime $(q\gtrsim1),$ for which the evaluation of the dispersion relation in the complex plane fails due to divergence of the required series. 
We can thus compute the critical velocity for absolute/convective instability for all confinement values, including the approach to the previously known result for the unconfined limit~\cite{huerre1992,brancher1997}. 

Similar to the results on the tanh profile reported previously by Healey~\cite{healey2009}, we obtain a range in which the effect of confinement is to increase the propagation velocity of the instability, thus rendering it potentially absolute in conditions where it would be convective without confinement. 
This leads to the surprising possibility of absolute instability without counterflow. 
However, this particular aspect is likely related to the particular model of point vortices since, without confinement, the results of Brancher \& Chomaz~\cite{brancher1997} for the model of Stuart vortices show that the propagation of the instability is slower with less concentrated vortices (see their Fig.~4). 
Still, the results of Ref.~\cite{brancher1997} are consistent with ours in that the secondary instability becomes absolutely unstable with less counterflow than the primary instability. 
This suggests that in a more realistic model the quantitative values will differ, but the overall picture 
can be expected to remain valid.

Assuming values of the mid-channel width $h$ that are not too small, the primary and secondary instabilities are, as expected, both absolute (1A2A) or both convective (1C2C) for sufficiently large or small values of the velocity ratio $R,$ respectively. 
However, the region of absolute/convective instability for the pairing instability cannot be determined unequivocally in the $(h,R)$-plane, because the threshold depends on the inter-vortex spacing $\lambda.$ 
This inter-vortex spacing physically corresponds to the wavelength selected by the primary instability, or by previous instances of the secondary instability, and is therefore left in the analysis as a free parameter to be determined by the dynamics. 
As a result, situations 1A2C or 1C2A are both possible depending on the values of the velocity ratio $R$ and wavelength $\lambda$ (Fig.~\ref{fig:ConvectiveAbsoluteBehavior}). 
Fixing $R$ sets the absolute/convective nature of the primary instability, but the pairing instability can generally still be absolute or convective depending on the wavelength $\lambda.$ 
For sufficiently small $\lambda,$ the secondary pairing instability becomes absolute with less counterflow than the primary instability. 
Increasing the wavelength $\lambda$, as it would occur through vortex pairings, eventually leads to the pairing instability becoming only convectively unstable. 

Cases in which the secondary pairing instability is convective have a predictable outcome: the row of vortices issuing from the saturation of the primary instability will be sensitive to incoming disturbances and exhibit irregular pairings as the vortices are advected downstream. 
In the common case of co-flowing shear layers~\cite{hoHuerr1984} the primary instability is also convective (1C2C) and the primary row of vortices typically also irregular. 
The situation will not change dramatically if the primary instability is absolute (1A2C), as in the case of the B\'enard-von K\'arm\'an street in the cylinder wake~\cite{mowlavi2016}; the secondary pairing instability might then still develop irregularly while advected on a regular row of vortices. 

Cases of absolute secondary instability are less straightforward, and should be generally studied on a case by case basis. 
A generic scenario that has been proposed is that of a sudden transition to complex behaviour~\cite{ChomazARFM2005}. 
This has been argued to be the case for free shear layers, for which the secondary instability is absolute before the primary instability in the absence of confinement~\cite{brancher1997}. 
However, for the present secondary instability, which is subharmonic, the increase in wavelength through successive pairings will also increase the effects of confinement, eventually leading to the instability becoming convective. 
In addition, since the primary instability is present for arbitrarily large wavelengths, its spatial saturation could 
lead to a row of vortices with a sufficiently large wavelength to be convectively unstable. 
These observations lead us to propose a mechanism for wavelength selection in confined shear layers with counterflow, which is consistent with numerical experiments of Ref.~\cite{Chomaz2004EJMBSecondaryAbsoluteAA}. 
According to this wavelength selection mechanism, spatially developing rows of vortices, which respect the global flow structure of mixing layers, are only possible with a confinement strong enough so that the pairing instability is convective. 
This could explain the lack of experimental realizations of spatial shear layers with counterflow, in particular the failed attempt of Ref.~\cite{humphreyLi1981tilting}. 
In addition, it provides a guideline for experimental and numerical search for shear layers with counterflow. 

In conclusion, we propose a rationale for explaining what are the conditions under which it can be possible to obtain spatial shear layers with counterflow.
Our results, however, being obtained in the highly idealized model of point vortices, are not expected to be quantitatively valid.
Still, experimental observations of a confined B\'enard-von K\'arm\'an vortex street~\cite{BonifaceEtalEPL2017} in the absence of advection have recently vindicated old predictions for the stability of these confined vortex streets based on von K\'arm\'an's point vortex model~\cite{Rosenhead1929}.
Also, our previous results reconciling K\'arm\'an's point vortex model with ubiquitous observations of vortex streets~\cite{mowlavi2016} further supports the utility of models of point vortices for studying hydrodynamic instabilites.

Obvious ways in which the present results could and should be extended involve asymmetric confinement and more realistic vortex street models, including but not limited to Stuart vortices with finite core size. 
The latter should lead to more accurate predictions for wavelength selection, and the details of the proposed mechanism could be then tested with numerical simulations. 
But ultimately, it should be the failure or success of experimental tests which shall eventually confirm or falsify our predictions. 
In this regard, soap films or an adaptation of the recent experiments~\cite{BonifaceEtalEPL2017} proving the stability of confined vortex streets seem promising,

\appendix
\section{Legendre transform method applied to K\'arm\'an's street of point vortices}\label{ap:karmanLegendre}

In this appendix, we present the application of the method developed in Section III A to a system with a more complicated dispersion relation than that of the confined single row. Specifically, we consider the dispersion relation of the unconfined and inviscid K\'arm\'an street of point vortices \citep{lamb1932,saffman1992,mowlavi2016} given by
\begin{equation}
\omega = \pm A +  s \sqrt{B^2-C^2},
\label{eq:KarmanStreetDispersionRelation}
\end{equation}
where $+A$ and $-A$ correspond to symmetrical and antisymmetrical modes respectively, $s = \pm 1$ gives two solution branches for each mode, and the coefficients $A$, $B$ and $C$ are expressed as
\begin{align}
A &= \frac{1}{2} k (2\pi-k) - \frac{\pi^2}{\cosh^2 p\pi}, \\
B &= \frac{\pi k \sinh p(\pi-k)}{\cosh p\pi} + \frac{\pi^2 \sinh pk}{\cosh^2 p\pi}, \\
C &= \frac{\pi^2 \cosh pk}{\cosh^2 p\pi} - \frac{\pi k \cosh p(\pi-k)}{\cosh p\pi},
\end{align}
\begin{figure}
\centering
\includegraphics[scale=0.6]{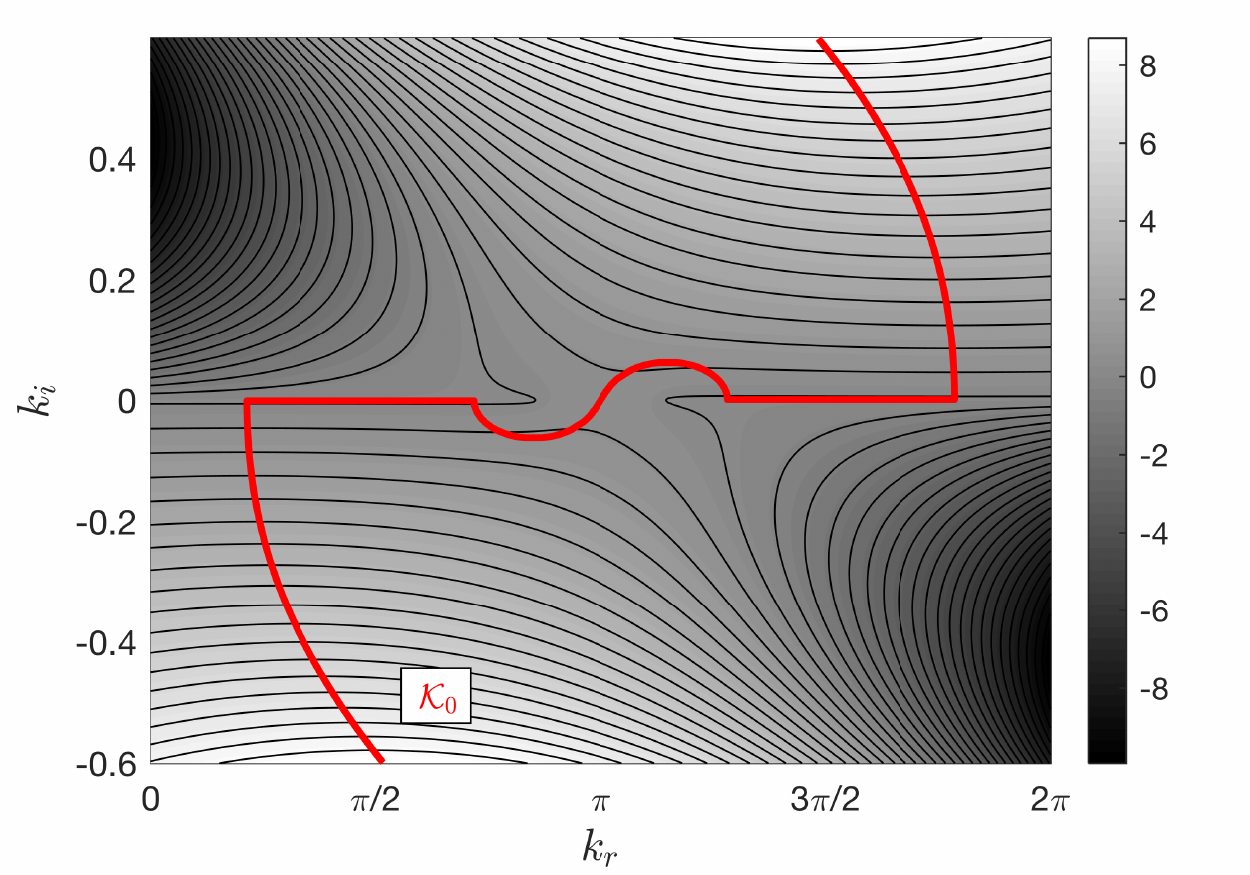}
\includegraphics[scale=0.6]{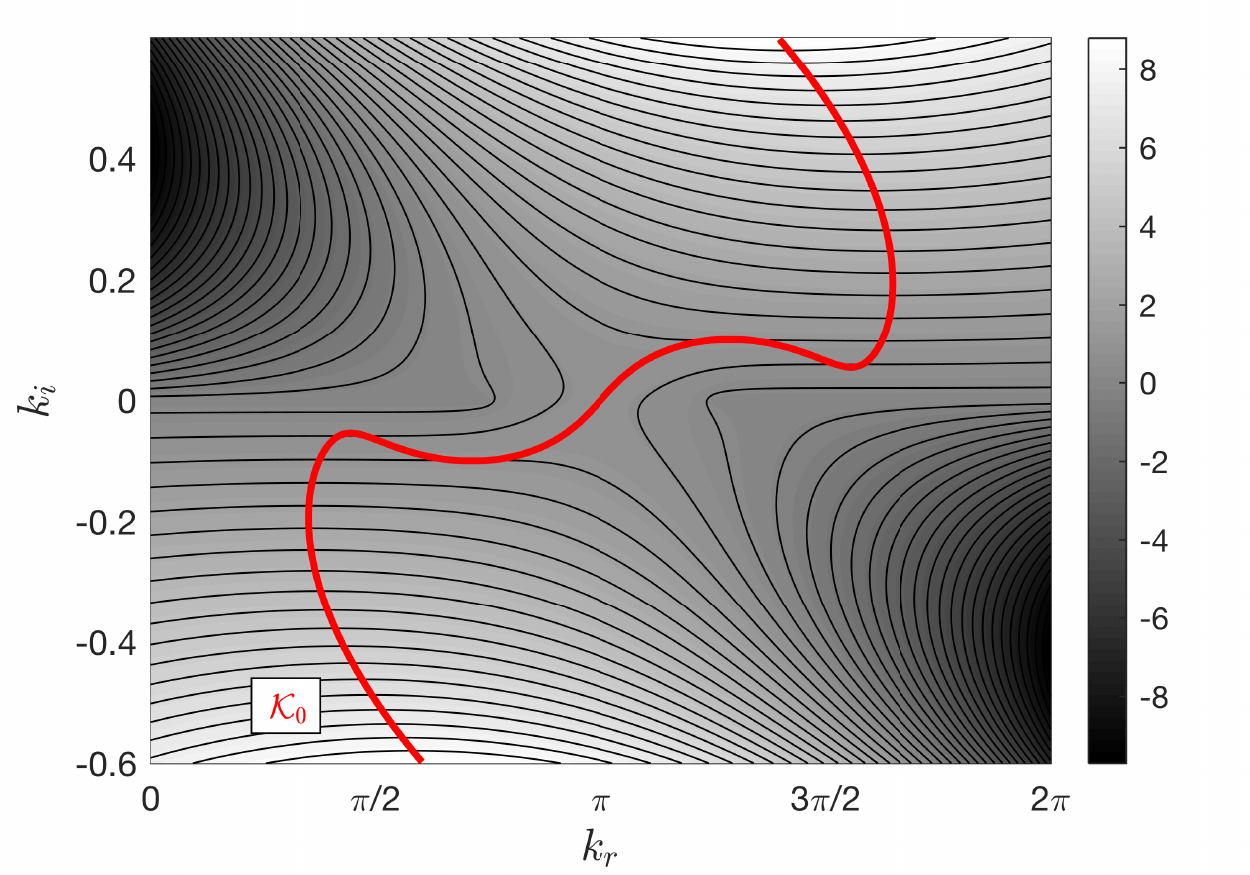}
\caption{Isocontours of $\omega_i$ in the complex $k$-plane for the dispersion relation of the K\'arm\'an street \eqref{eq:KarmanStreetDispersionRelation} with $p = 0.3$ (left) and $p = 0.316$ (right). The red line represents the locus $\mathcal{K}_0$ of absolute wavenumbers defined by (19). Note that in both cases, $\mathcal{K}_0$ contains multiple values of $k_r$ for some intervals of $k_i$.}
\label{fig:KarmanStreet}
\end{figure}%
\begin{figure}
\centering
\includegraphics[scale=0.6]{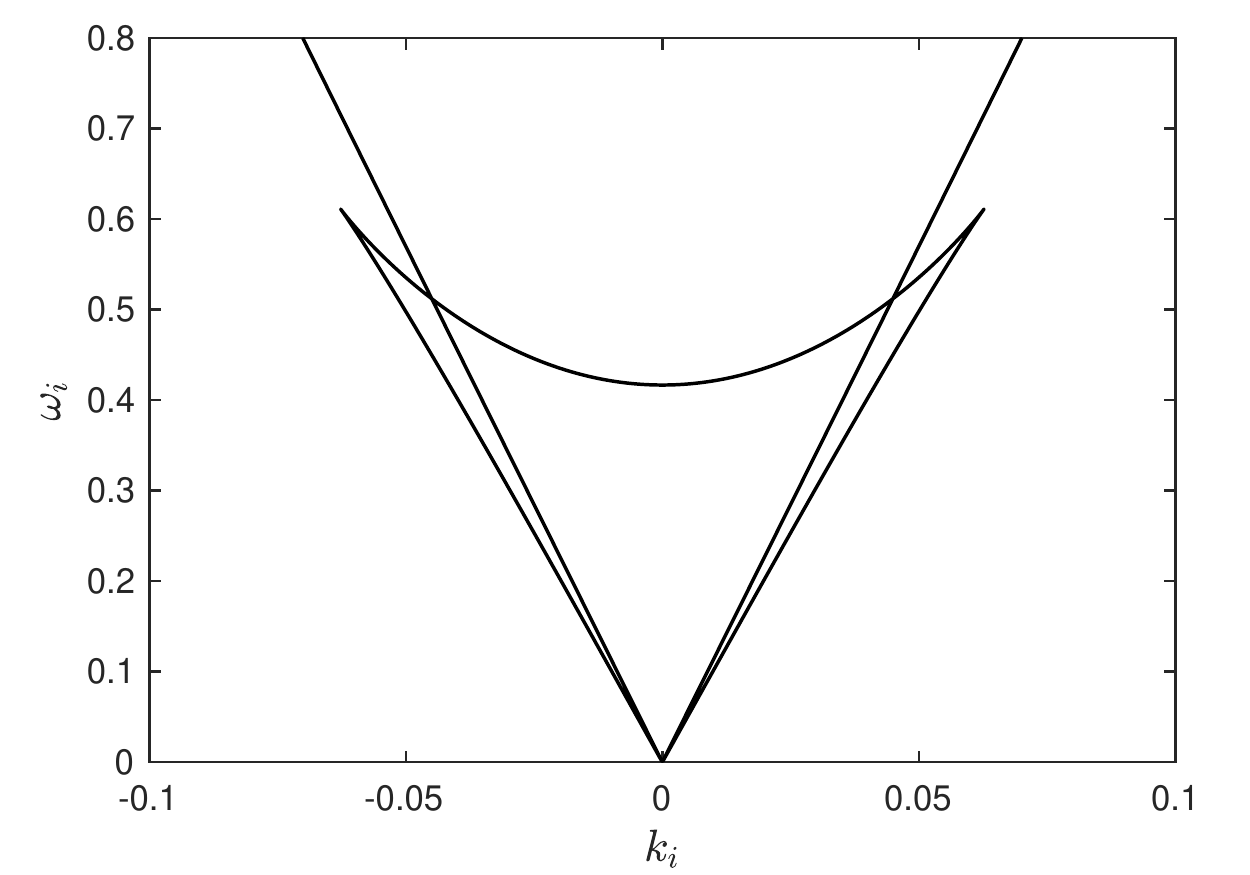}
\includegraphics[scale=0.6]{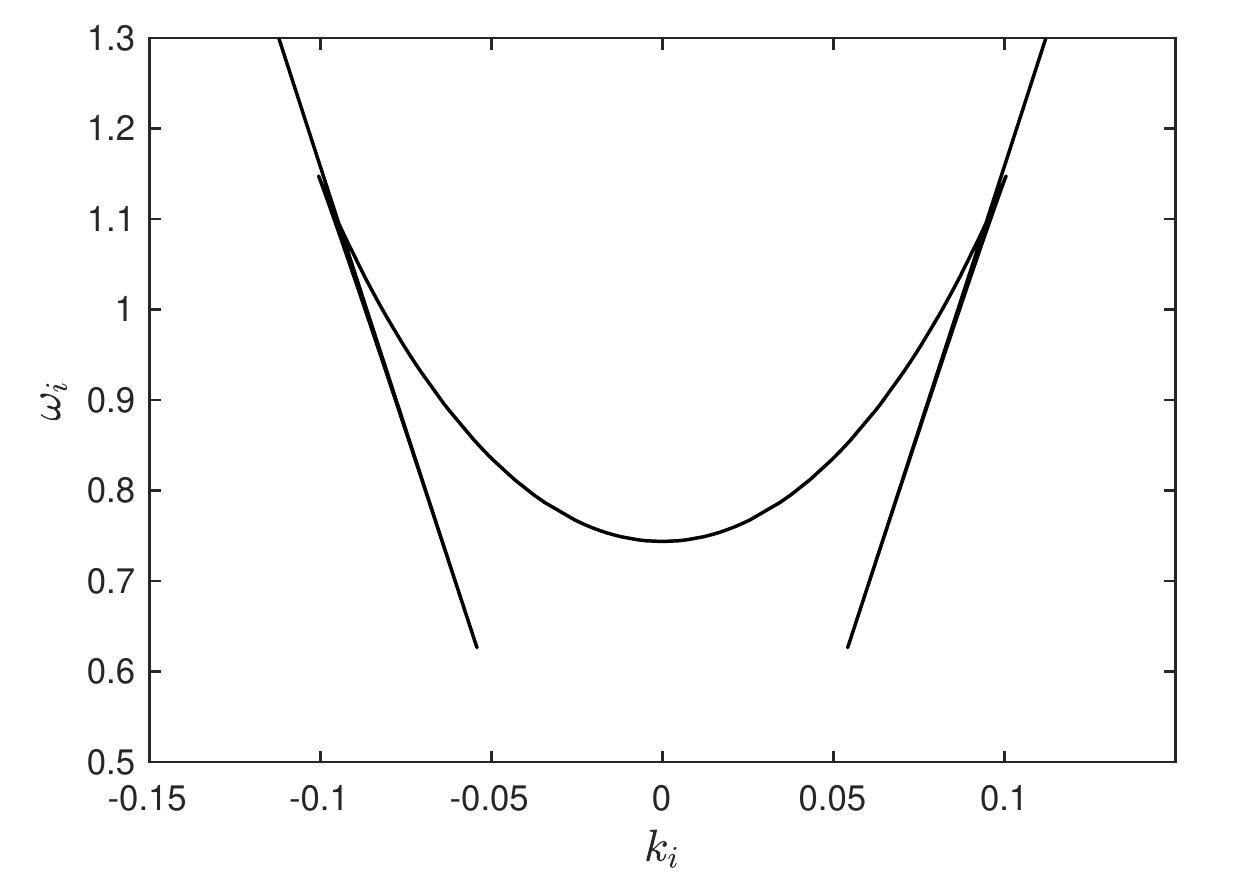}
\caption{Imaginary part of the frequency along the locus $\mathcal{K}_0$ of absolute wavenumbers, expressed as a multivalued function $\omega_i(k_{0r}^v(k_i),k_i)$ of $k_i$ for $p = 0.3$ (left) and $p = 0.316$ (right).}
\label{fig:KarmanStreet_wiki}
\end{figure}%
where $p$ is the ratio of the vertical distance between the two rows of vortices to the horizontal distance between consecutive vortices in the same row. 
The wavenumber $k$ of the perturbation is nondimensionalized by the inverse of the distance between consecutive vortices. 
In Figure \ref{fig:KarmanStreet}, we plot the contour levels of $\omega_i(k)$ from \eqref{eq:KarmanStreetDispersionRelation} for $p = 0.3$ (left) and $p = 0.316$ (right), together with the locus $\mathcal{K}_0$ of absolute wavenumbers defined by Eq.\eqref{eq:LocusAbsoluteWavenumbers}. 
Unlike the earlier case of the single row of vortices, here $\mathcal{K}_0$ contains multiple values of $k_r$ for some intervals of $k_i$. 
We now apply the geometrical technique of Section \ref{sec:LegendreGeoConstruction} for obtaining the growth rate of an impulse response wave packet along spatiotemporal rays $x/t = v$. 
First, we plot in Figure \ref{fig:KarmanStreet_wiki} the curve $\omega_i(k_{0r}^v(k_i),k_i)$ for $p = 0.3$ (left) and $p = 0.316$ (right), which represents the imaginary part of the frequency along $\mathcal{K}_0$. 
Note that, as always in these plots, $\omega_i$ intersects the axis $k_i=0$ at the maximum of the temporal growth rate. 
For the present dispersion relation, $\omega_i(k\in\mathcal{K}_0)$ is a multivalued function of $k_i$. 
At the point where $\mathcal{K}_0$ becomes parallel to the $k_r$-axis and $\omega_i$ becomes multivalued, Eq. \eqref{eq:variesOnlyWithKi} prevents $\omega_i$ from becoming vertical, and it generically develops a cusp instead, as shown in Figure \ref{fig:KarmanStreet_wiki}(right). 
Another possibility for this conservative system can be seen in Figure \ref{fig:KarmanStreet}(left), in which $\mathcal{K}_0$ coincides with the $k_r$-axis for a finite distance and the tangent to $\omega_i(k_{0r}^v(k_i),k_i)$ has a discontinuity when touching the origin in Figure \ref{fig:KarmanStreet_wiki}(left). 
Nevertheless, the monotonous behavior of $\partial\omega_i/\partial k_i = v$ along $\mathcal{K}_0$ ensures that every point on the locus corresponds to a unique value of $v$. 
This allows for the unambiguous determination of the growth rate of the impulse response wave packet based on \eqref{eq:LegendreTransform} and the geometrical construction detailed in Section \ref{sec:LegendreGeoConstruction}. 
The resulting curve $\sigma(v)$ is shown in Figure \ref{fig:KarmanStreet_sigma} for $p = 0.3$ (left) and $p = 0.316$ (right). 

\begin{figure}
\centering
\includegraphics[scale=0.6]{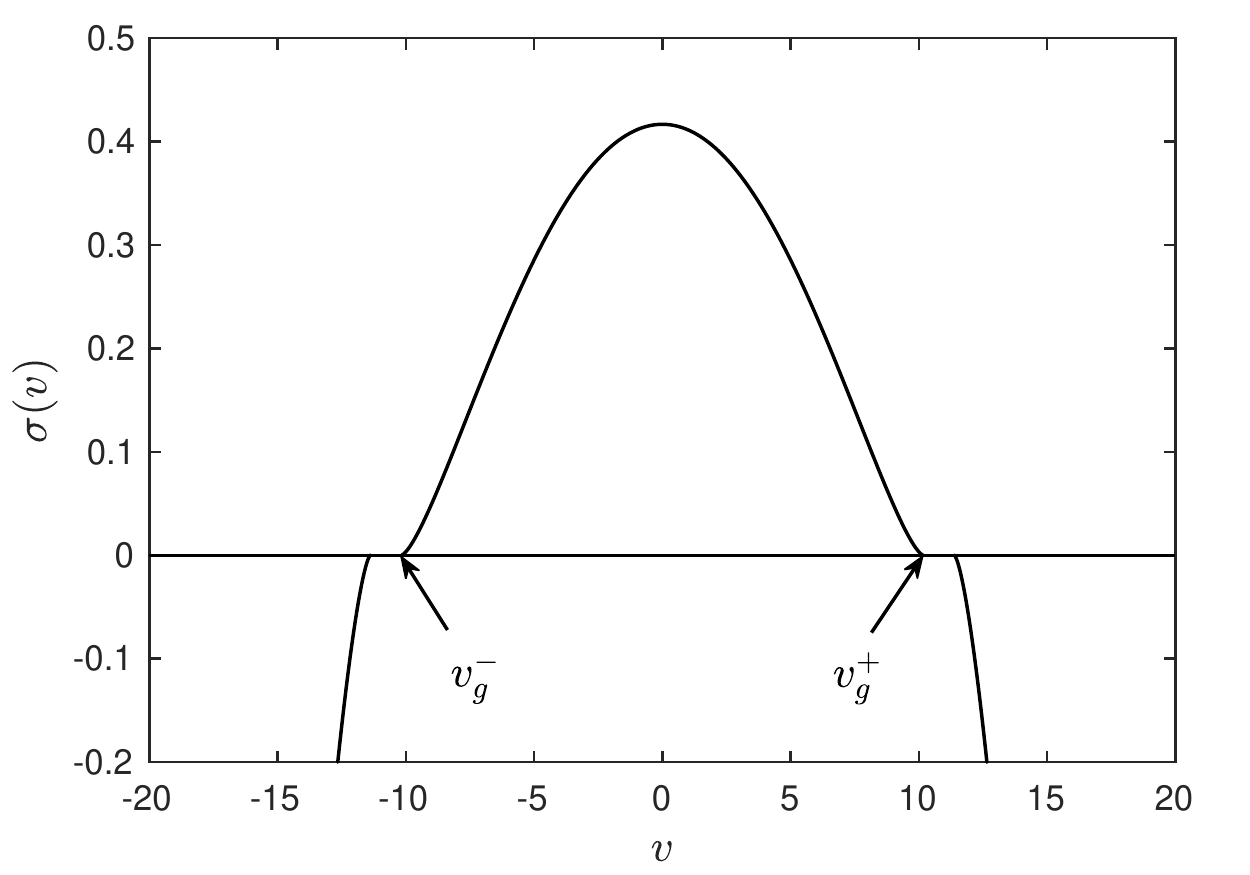}
\includegraphics[scale=0.6]{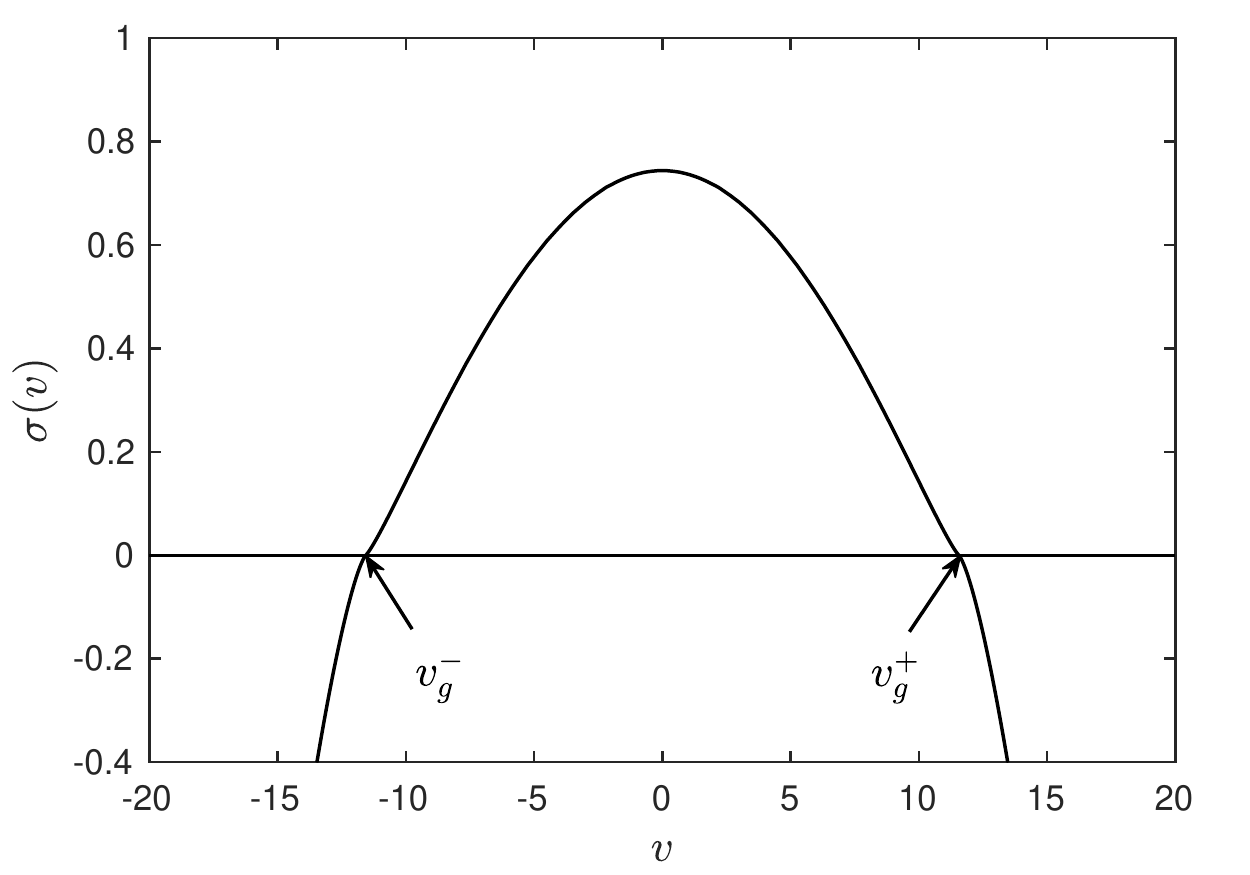}
\caption{Growth rate $\sigma(v)$ of the impulse response wave packet along spatio-temporal rays $x/t = v$ for $p = 0.3$ (left) and $p = 0.316$ (right). The streamwise extent of the wave packet is given by the leading- and trailing-edge velocities $v_g^\pm$ such that $v_g^\pm = 0$.}
\label{fig:KarmanStreet_sigma}
\end{figure}%

\section{Determination of the primary instability boundary}\label{ap:ExplainR2C}

The function $\bar{R}_{2C}(h)$, which delimits the region in which the $R_{c2,\lambda}(h)$ family describes different A/C regions for secondary stability, can be obtained as illustrated in the inset of Fig.~\ref{fig:ConvectiveAbsoluteBehavior}. 
Take a value of $\lambda=\bar{\lambda}$ slightly larger than $4\pi$ (say for example $\bar{\lambda}=4.4\pi$ shown in one of the continuous lines of the inset of Fig.~\ref{fig:ConvectiveAbsoluteBehavior}, corresponding to $\bar{k}=2\pi/\bar{\lambda}=1/2.2$) and follow its corresponding $R_{c2,\bar{\lambda}}$ curve while decreasing $h$ until the value $\bar{h}$ for which the corresponding $\bar{k}$ is critical for the $\mathrm{tanh}$ profile, i.e., the $\bar{h}$ such that $k_c(\bar{h})=\bar{k}.$ 
This determines $\bar{R}_{2C}(h)$ for $h=\bar{h}$ as $\bar{R}_{2C}(\bar{h})=R_{c2,\bar{\lambda}}(h=\bar{h})$. 
Below this $\bar{h}$ the wavelength $\bar{\lambda}$ cannot result from the primary instability since it is stable for $\bar{k}$. 
Indeed, for an $h<\bar{h}$ the wavelength from the primary instability must be a $\lambda>\bar{\lambda}$, which would select a critical curve $R_{c2,\lambda}$ to the right of $R_{c2,\bar{\lambda}}$ that will not extend below $\bar{R}_{2C}(h).$ 
Conversely, if we fix $h$ and take a value of $R$ below the $\bar{R}_{2C}(h)$ curve, the primary instability will necessarily select a wavelength $\lambda$ such that $R_{c2,\lambda}$ lies above $\bar{R}_{2C}(h),$ and the secondary instability will be convective.


\input{main.bbl}

\end{document}